\documentclass[letterpaper,11pt]{article}
\usepackage{amsmath,amssymb,amsthm}
\usepackage[dvips]{graphicx}
\usepackage[letterpaper,nohead,left=1in,right=1in,top=1.2in]{geometry}
\begin{document}

\newtheorem{theo}{Theorem}[section]
\newtheorem{coro}[theo]{Corollary}
\newtheorem{defi}[theo]{Definition}
\newtheorem{exam}[theo]{Example}
\newtheorem{lemm}[theo]{Lemma}
\newtheorem{conj}{Conjecture}
\newtheorem{prop}[theo]{Proposition}
\newtheorem{prope}[theo]{Property}

\def\R{\mathbb{R}}
\def\Reals#1{\mathbb{R}^{#1}}

\def\notyetfound#1{\mbox{{\rm NOT-YET-FOUND}}\left(#1  \right)}
\def\found#1{\mbox{{\rm FOUND}}\left(#1  \right)}
\def\name#1{\mbox{{\sf name}}\left(#1  \right)}
\def\tail#1{\mbox{{\sf tail}}\left(#1  \right)}
\def\height#1{\mbox{{\sf height}}\left(#1  \right)}
\def\level#1{\mbox{{\sf level}}\left(#1  \right)}
\def\head#1{\mbox{{\sf head}}\left(#1  \right)}
\def\eend#1#2{\mbox{{\sf end}}_{#1}\left(#2  \right)}
\def\EventTrue#1{\left[  #1 \right]}
\def\inser#1#2#3{\mbox{{\sf insert}}_{#1}\left(#2,#3 \right)}
\def\setof#1{\left\{{\let\st\colon #1 }\right\}}
\def\Zdn{\mathbb{Z}^d_n} \def\Ed{\mathbb{E}^d} \def\BBO{\mathbb{B}} \def\Real{\mathbb{R}}
\def\NTd{{\sf NT}$^d$ }\def\mbbJ{\mathbb{J}}\def\mbbO{\mathbb{O}}\def\mbbF{\mathbb{F}}
\def\ssigma{\mathbf{\sigma}}
\def\aa{\mathbf{a}} \def\bb{\mathbf{b}} \def\cc{\mathbf{c}}
\def\ee{\mathbf{e}}
\def\00{\mathbf{0}} \def\11{\mathbf{1}}
\def\uu{\mathbf{u}} \def\vv{\mathbf{v}} \def\xx{\mathbf{x}} \def\yy{\mathbf{y}}
\def\AA{\mathbf{A}} \def\BB{\mathbf{B}} \def\LL{\mathbf{L}} \def\UU{\mathbf{U}}
\def\pp{\mathbf{p}} \def\qq{\mathbf{q}} \def\rr{\mathbf{r}} \def\ss{\mathbf{s}}
\def\tt{\mathbf{t}} \def\xx{\mathbf{x}} \def\yy{\mathbf{y}} \def\ZZ{\mathbb{Z}}
\def\zz{\mathbf{z}} \def\LL{\mathbf{L}} \def\RR{\mathbf{R}} \def\MM{\mathbf{L}}
\def\NN{\mathbf{R}} \def\calG{\mathcal{G}} \def\calD{\mathcal{D}} \def\calA{\mathcal{A}} \def\calI{\mathcal{I}}
\def\calS{\mathcal{S}} \def\calP{\mathcal{P}} \def\calC{\mathcal{C}} \def\size#1{Size$[#1]}
\def\bbE{\mathbb{E}} \def\ZZ#1#2{\mathbb{Z}^{#1}_{#2}} \def\ww{\mathbf{w}} \def\calL{\mathcal{L}}
\def\calF{\mathcal{F}} \def\calT{\mathcal{T}} \def\calE{\mathcal{E}} \def\calP{\mathcal{P}} \def\calR{\mathcal{R}}
\def\calB{\mathcal{B}} \def\22{\mathbf{2}}
\def\calK{\mathcal{K}}

\thispagestyle{empty}
\title{\LARGE Paths Beyond Local Search: A Nearly \vspace{0.1cm}Tight Bound \\for
Randomized Fixed-Point\vspace{0.7cm}
Computation\thanks{This research is supported mostly by the NSF ITR grant CCR-0325630.}}
\author{
    Xi Chen\thanks{Part of this work was done while visiting Computer Science Department at
Boston University.
In part supported by the Chinese National Key Foundation R \& D Plan
  (2003CB317807, 2004CB318108), the National Natural Science
  Foundation of China Grant 60553001 and the
  National Basic Research Program of China Grant (2007CB807900, 2007CB807901).
} \\ {\small Department of Computer Science} \\ {\small Tsinghua University} \\
    \and
    Shang-Hua Teng\thanks{Part of this work was done while visiting Tsinghua University and Microsoft Research Asia Lab.} \\ {\small Department of Computer Science} \\ {\small Boston University} \\
} \date{} \maketitle

\begin{abstract}

In 1983, Aldous proved that randomization can speedup local search.
For example, it reduces
  the query complexity of local search over $[1:n]^d$ from
  $\Theta (n^{d-1})$ to $O (d^{1/2}n^{d/2})$.
It remains open whether randomization helps fixed-point computation.
Inspired by this open problem and recent advances on
  equilibrium computation,
  we have been fascinated by the following question:

\begin{center}
{\em  Is a fixed-point or an equilibrium fundamentally harder to find than a local
  optimum? \vspace{0.05cm}}
\end{center}

In this paper, we give a nearly-tight bound of $\left(\Omega
  \left(n\right) \right)^{d-1}$
  on the randomized query
  complexity for computing a fixed point
  of a discrete Brouwer function over  $[1:n]^d$.
Since the randomized query complexity of
  global optimization over $[1:n]^d$ is $\Theta (n^{d})$,
  the randomized query model over $[1:n]^d$
  strictly separates these three important search problems:\vspace{0.1cm}

\begin{center}
{\em Global optimization is harder than
  fixed-point computation, and \\
    fixed-point computation is harder than local search}.\vspace{0.1cm}
\end{center}

Our result indeed demonstrates that randomization does not help much in
  fixed-point computation in the query model; the deterministic
  complexity of this problem is $\Theta (n^{d-1})$.






\end{abstract}
\thispagestyle{empty}
\newpage
\thispagestyle{empty}
\section*{Prologue}

\begin{quotation}
\noindent
\begin{itemize}
\item [Scene 1:]
{\em
On the first day of your new job, your boss teaches you the Simplex
  Algorithm with the Steepest-Edge Pivoting Rule.
You quickly master the steps of the algorithm. So
  she gives you a large linear program that simulates a new
  business model.

``I am going to a convention in Hawaii for ten days.
Could you work on the program starting with
  this initial vector $\xx_{0}$?''
  she asks. ``The solution will be a vector that you cannot improve upon.
  Email it to me when you are done''

So she leaves for beautiful Hawaii and you begin your iterative path-following
  search.
Ten days later, she comes back, relaxed,
  right as you finish computing $\xx_{1000000}$!

``I haven't found the solution yet,'' you report, ``but I have followed the
  steepest-edges a million steps and get $\xx_{1000000}$.''

She takes the objective vector $\cc $ and quickly computes
  $\cc^{T}\xx_{1000000}/\cc^{T}\xx_{0}$, and it is $1.10$.

``You find a vector that is 10 percent better than what we had
  initially,'' she says  cheerfully.  ``Good job!''

The next day, you get a ten-percent raise.
}

\vspace{0.2in}

\item [Scene 2:]
{\em
On the first day of your new job, your boss teaches you the Lemke-Howson
  algorithm for finding a Nash equilibrium of a two-player game.
You quickly master the steps of the algorithm.
So he gives you a large two-player game that
  models a two-group exchange market.

``I am going to a convention in Hawaii for ten days.
Could you work on this two-player game?'' he asks.
``Here is an initial strategy-profile,''  he gives you $(\xx_{0},\yy_{0})$,
``and the strategy-profile that Lemke-Howson halts on is a
  Nash equilibrium. Email it to me when you are done.''

So he leaves for beautiful Hawaii and you begin your iterative path-following
  search.
Ten days later, he comes back, relaxed,
  right as you finish computing $(\xx_{1000000},\yy_{1000000})$!

``I haven't found the solution yet,'' you report, ``but I have followed the
  Lemke-Howson path a million steps and get $(\xx_{1000000},\yy_{1000000})$.''

He looks at $(\xx_{1000000},\yy_{1000000})$ for a while and then frowns,
  just slightly.

``Hmmmm, no equilibrium in a million steps!'' he says. ``Well, good job and thanks.''

The next day, you still have your job but get no raise.
}
\end{itemize}
\end{quotation}

\newpage
\setcounter{page}{1}
\section{Introduction}

The Simplex Algorithm \cite{Dantzig}
  is an example of an implementation of local
  search\footnote{Note that in linear programming,  each local optimum is also a global optimum.}
  and finding a Nash equilibrium \cite{NAS50}
  is an example of fixed-point computation (FPC).
A general approach for local search is Iterative
Improvement.
Steepest-Descent is its most popular example.
It follows a path in the feasible space,
  a path along which the objective values are monotonically improving.
The end of the path is a local optimum.
Like Iterative Improvement, many algorithms for
  FPC, such as the
  Lemke-Howson algorithm \cite{LemkeHowson} and the constructive proof
   of Sperner's Lemma \cite{Sperner}, also follow a path whose
  endpoint is an equilibrium or a fixed-point.
But unlike a path in local search, a path in
  FPC does not have an obvious ``locally
  computable'' monotonic\footnote{Each path has a ``globally
computable''
  monotonic measure, the number of hops from the
  start of the path to a node.} measure-of-progress.
Moreover, path following in FPC
  from an arbitrary point could lead to a cycle while
  the union of paths in Iterative Improvement is acyclic.

\begin{center}
{\em Do these structural differences have any algorithmic implication? }
\end{center}

There have been increasing evidence, beyond the stories of our prologue,
  that local search and FPC are very different.
First, Aldous \cite{ALD}
  showed that randomization can speedup local search (more discussion below).
His method crucially utilizes the monotonicity discussed above.
It remains open whether randomization helps FPC.
Second, polynomial-time path-following-like algorithms
  have been developed for
  some non-trivial classes of local search problems.
These algorithms include the interior-point algorithm
  for linear and convex programming \cite{Karmarkar,InteriorBible} and
  edge-insertion algorithms for geometric
  optimization \cite{edels}.
However, popular fixed-point problems, such as the computation
  of a Nash or a market equilibrium \cite{ArrowDebreu}
  might be hard for polynomial time \cite{DAS05,CHE06,CSVY}.
Other than those that can be solved by convex programming,
  we haven't yet discovered a significantly non-trivial class of
  equilibrium problems that are solvable in polynomial-time.
Third, an approximate local optimum for every PLS (Polynomial Local
  Search) problem
  can be found in fully-polynomial time
\cite{OrlinPunnenSchulz}.
In contrast, although a faster randomized algorithm was found for approximating
  Nash equilibria \cite{LIP03},
   finding an approximate Nash equilibrium
   in fully-polynomial time is computationally equivalent to
   finding an exact Nash equilibrium in polynomial time \cite{CDT06}.
We face the same challenge in approximating market equilibria \cite{HuangTeng}.
Fourth, although they all have exponential worst-case complexity
  \cite{SavanivonStengel,KleeMinty},
  the smoothed complexity of the Simplex Algorithm and Lemke-Howson
  Algorithm (or Scarf's market equilibrium algorithm \cite{ScarfBook}) might be drastically
  different \cite{SpielmanTengSimplex,CDT06,HuangTeng}.
This evidence inspires us to ask:

\begin{center}
{\em  Is fixed-point computation fundamentally harder than local
  search? }
\end{center}




To investigate this question, we consider
  the complexity of these two search problems defined over
  $\mathbb{Z}^d_n=[1:n]^d$.
For fixed-points, we are given a function $F: \Zdn \rightarrow
  \Zdn$ that satisfies  Brouwer's condition \cite{Brouwer} ---
   a set of continuity and boundary conditions (see  Section 2) ---
   that guarantees the existence of a fixed-point.
Recall that  a vector $\vv \in \Zdn $  is a fixed-point of $F$ if $F (\vv ) = \vv $.
The {\em FPC problem}
is to find a   fixed-point of  $F$.
For local optima, we are given a function $h: \Zdn
  \rightarrow \Real$.
The {\em local search problem}
  is to find a local optimum of $h$, for example,
  a vector $\xx \in \Zdn $ such that $h (\xx ) \geq h (\yy )$,
  $\forall \yy $ with $||\xx-\yy||_{1} \leq 1$.

For both problems, we consider the query complexity in the query
  model:
The algorithm can only access $F$ and $h$, respectively, by asking queries of the form:
  ``What is $F(\xx )$?'' and ``What is $h(\xx )$?''.
The complexity is measured by the number of queries needed to
  find~a~solution.

\def\RQ#1{\text{\sf RQ}_{\text{\sf #1}}^d(n)}

There are some similarities between FPC and
  local search over $\Zdn$.
For both, divide-and-conquer
  has positive but limited success:
Both problems can be solved by $O (n^{d-1})$ queries \cite{XX05}.
An alternative approach to solve both problems is
  path-following.
When following a short path, it can be
  faster than divide-and-conquer.
But for both problems, long and winding paths are the cause of
  inefficiency.

However, there is one prominent difference between a path to a local
  optimum and a path to a fixed point.
The values of $h$ along a path to a local optimum are
  monotonic, serving as a measure-of-progress along the path.
Aldous \cite{ALD} used this fact in a randomized algorithm:
  Randomly query $d^{1/2}n^{d/2}$ points in $\Zdn$;
  let $\ss$ be the sample point with the largest $h$ value;
follow a path starting at $\ss$.
If a path to a local optimum is long, say much longer than
$d^{1/2}n^{d/2}$,
  then with high probability, the random samples intersect
  the path and partition it into
  sub-paths, each with expected length $O(d^{1/2}n^{d/2})$.
As $\ss$ has the largest $h$ value,
  its sub-path is the last sub-path of a potentially long path, and
  we expect its length to be $O(d^{1/2}n^{d/2})$.
So with randomization, Aldous reduced the expected query
  complexity to $O (d^{1/2}n^{d/2})$.\vspace{0.02cm}

But it remains open
  whether randomization can reduce the query complexity
  of FPC over $\Zdn $.
The lack of a measure-of-progress along a path makes it impossible for us
   to directly use Aldous' idea.

\subsection*{Our Main Result}

The state of our knowledge suggests that
   FPC might be significantly harder than
   local search, at least in the randomized query model.
We have formulated a concrete conjecture stating that
  an expected number of $\left(\Omega \left(n\right) \right)^{d-1}$
  queries are needed in randomized FPC over $\Zdn$.

As the main technical result of this paper, we prove
  that an expected number of
   $\left(\Omega \left(n\right) \right)^{d-1}$ queries
   are indeed needed.
Our lower bound is essentially
   tight\footnote{The constant in $\Omega$ in our lower
bound depends exponentially on $d$. See Theorem \ref{thm:FixedPoint}.},
  since the deterministic divide-and-conquer algorithm in
  \cite{XX05} can find
  a fixed point by querying $O (n^{d-1} )$ vectors.
In contrast to Aldous's result~\cite{ALD},
  our result demonstrates that randomization does not help much in
  FPC in the query model.
It shows that,
  in the randomized query model over $\Zdn$,
  a fixed-point is strictly harder to find than a local
  opti\-mum!
The significant gap between these two
   problems is revealed only in randomized computation.
In the deterministic framework, both have query
  complexity $\Theta  (n^{d-1})$.\vspace{0.01cm}

One can show that the randomized query complexity
  for finding a global optimum over $\Zdn$ is $\Theta(n^{d})$.
So, the randomized query model over $\Zdn $ strictly separates
  these three important~search~problems:

\begin{center}
{\em Global optimization is harder than
  fixed-point computation, and \\
    fixed-point computation is harder than local search}.
\end{center}


We anticipate that a similar gap can be obtained in the quantum
  query  model.

\subsection*{Related Work and Technical Contributions}

Our work is  also
  inspired by the lower bound results of Aaronson \cite{AAR},
  Santha and Szegedy \cite{SAN}, Zhang \cite{ZHANG}, and
  Sun and Yao \cite{SUN} on the randomized and quantum query complexity
  of local search over $\Zdn$.\vspace{0.01cm}

In this paper, we introduce several new techniques to study the complexity of
  FPC.
Instrumental to our analysis, we develop a method to generate
  hard-to-find random long paths in the grid graph over $\Zdn $.
To achieve our nearly-tight lower bound, these paths must be much longer
  than the random paths constructed in \cite{ZHANG,SUN} for local search.
Our paths has expected length $\left(\Theta (n) \right)^{d-1}$ while
  those random paths for local search have length $\Theta  (n^{d/2})$.
We also develop new techniques for unknoting a self-intersecting path
  and for realizing a path with a Brouwer function.
These techniques might be useful on their own in the future
  algorithmic and complexity-theoretic studies of FPC and its applications.\vspace{0.01cm}

There are several earlier work on the
  query complexity of FPC.
Hirsch, Papadimitriou and Vavasis \cite{HPV} considered
  the deterministic query complexity of FPC.
They proved a tight $\Theta(n)$ bound for
  ${\mathbb{Z}^2_n}$ and an $\Omega
  (n^{d-2})$ lower bound for $\Zdn$.
Subsequently, Chen and Deng \cite{XX05}
  improved this bound to $\Theta(n^{d-1})$ for $\Zdn$.
Recently, Friedl, Ivanyos, Santha, and Verhoeven \cite{FISV}
  gave a $\Omega (n^{1/4})$-lower bound on the randomized
  query-complexity of the 2D Sperner problem.
Our method for unkonting self-intersecting paths
  can be viewed as an extension of the 2D technique of \cite{XX06}
  to high dimensions.

\subsection*{Paper Organization}

In Section 2, we introduce three high-dimensional search problems.
In Section 3, we reduce  one of them, called {\sf End-of-a-String},
  to fixed-point computation over $\Zdn$.
In Section 4, we give a nearly tight bound on the randomized
  query complexity of {\sf End-of-a-String}.
Together with the reduction in Section 3, we obtain our main
  result on fixed-point computation.

\section{Three High-Dimensional Search Problems}\label{sec:}

We will define three search problems.
The first one concerns FPC.
We introduce the last two to help the
  study of the first one.
Below, let   $\mathbb{E}^d=\{\pm \ee_1,\pm \ee_{2},...,\pm \ee_{d}\}$
  be the set of {\em principle unit-vectors} in $d$-dimensions.
Let $\|\cdot\|$ denote $\|\cdot\|_\infty$.
For two vectors\footnote{We will use bold lower-case
  Roman letters such as $\xx $, $\aa$, $\bb_{j}$ to denote vectors. 
Whenever a vector, say $\aa\in\Reals{n}$ is
  present, its components will be denoted by 
  lower-case Roman letters with subscripts, such as $a_{1},\dotsc ,a_{n}$.
So entries of $\bb_{j}$ are $(b_{j,1},\dotsc ,b_{j,n})$.
} $\uu\not=\vv$ in $\mathbb{Z}^d$,
  we say $\uu<\vv$ lexicographically if $u_i<v_i$ and
  $u_j=v_j$ for all $1\le j<i$, for some $i$.\vspace{0.01cm}

For each of the three search problems, we will define its mathematical structure,
  a query model for accessing this structure,
  the search problem itself, and its query complexity.

\subsection{Discrete Brouwer Fixed-Points}\label{}

Recall that a vector $\vv\in\Zdn $ is a {\em fixed-point} of
  a function $F$ from $\Zdn$ to $\Zdn$ if $F (\vv ) = \vv$.
A function $f: \ZZ{d}{n}\rightarrow \{\00\}\cup\bbE^d$ is
  {\em bounded} if $f(\xx)+\xx\in \ZZ{d}{n}$ for all $\xx\in
  \ZZ{d}{n}$; $\vv\in \Zdn$ is a {\em zero point} of $f$
  if $f (\vv ) =\00$.
Clearly, if $F (\xx ) = \xx + f (\xx)$
 for all $\xx \in\Zdn$,
  then $\vv $ is a fixed point of $F$ iff $\vv $ is a zero point of~$f$.

\begin{defi}[\mbox{\rm Direction Preserving Functions}]
A function $f$ from $S$ to $\{\hspace{0.03cm}\00\hspace{0.03cm}\}\cup \bbE^d$
  where $S\subset\mathbb{Z}^d$ is \mbox{\rm direction-preserving}
   if $\|f(\rr_{1})-f(\rr_{2})\|\le 1$  for all pairs
   $\rr_{1},\rr_{2}\in S$ such that $\|\rr_{1}-\rr_{2}\|\le  1$.
\end{defi}

Following the discrete fixed-point theorem
  of \cite{IIM}, we have:
For every function $f: \ZZ{d}{n}\rightarrow\{\00\}\cup\bbE^d$,
  if $f$ is both bounded and direction-preserving, then
  there exists $\vv\in \ZZ{d}{n}$ such that $f(\vv)=\00$.
We refer to a bounded and direction-preserving function
  $f$ over $\Zdn $ as a {\em Discrete Brouwer function}
  or simply a {\em Brouwer function} over $\Zdn$.
In the query model, one can only access $f$
  by asking queries of the form: ``What is $f(\rr)$?''
  for a query point $\rr\in \ZZ{d}{n}$.\vspace{0.02cm}

The FPC problem $\text{\sf ZP}^d$
  that we will study is as follows:
{\em Given a Brouwer function $f$ from $\ZZ{d}{n}$ to
    $\{\00\}\cup\bbE^d$ in the query model, find a zero point
    of $f$.}
Let $\text{\sf RQ}_{\text{\sf ZP}}(f)$ denote the
  expected number of queries needed by the best randomized algorithm
  to find\footnote{One can also change ``to find'' to ``to find, with high probability''.}
   a zero point of $f$.
We let
\[
\text{\sf RQ}_{\text{\sf ZP}}^d(n) = \max_{f:\ \text{Brouwer
function over $\Zdn $}} \big\{\hspace{0.04cm}\text{\sf RQ}_{\text{\sf ZP}}(f)\hspace{0.04cm}\big\},\vspace{-0.1cm}
\]
be the {\em randomized query complexity} for solving $\text{\sf ZP}^{d}$.
In this paper, we will prove:

\begin{theo}[\mbox{\rm Randomized Query Complexity of Fixed Points}] \label{thm:FixedPoint}
\hspace{-0.13cm}There is a constant $c$ such that for all sufficiently large~$n$,
\[\text{\sf RQ}_{\text{\sf ZP}}^d(n)\ge \left( \frac{n}{c^{d}}\right)^{d-1}.\]
\end{theo}

In contrast, the deterministic query complexity for
  solving $\text{\sf ZP}^{d}$ is at most $7n^{d-1}$ \mbox{\rm \cite{XX05}}.
The Brouwer fixed point problem defined here is computationally equivalent
  to the fixed problems defined in \cite{HPV,DAS05,CDT06}.
Thus, our result carries over to these FPC problems.

\subsection{End-of-a-Path in Grid-PPAD Graphs}\label{}

The mathematical structure for this search problem is a directed
  graph $G= (V,E)$.
A vertex $v\in V$ satisfies {\em Euler's condition}
  if $\Delta_{I} (v) = \Delta_{O} (v)$ where
  $\Delta_{I} (v)$ and $\Delta_{O} (v)$ are the in-degree and
   the out-degree of $v$.
We start with the following definition motivated by
   Papadimitriou's \textbf{PPAD} class \cite{PAP94}.

\begin{defi}[\mbox{\rm Generalized PPAD Graphs}]
A directed graph $G=(V,E)$ is a {\rm generalized PPAD graph} if (1)
  there exists exactly one vertex $v_S\in V$ with $\Delta_{O} (v_{S})
  =    \Delta_{I} (v_{S})+1$
  and exactly one vertex $v_T\in V$
  with $\Delta_{I} (v_{T}) = \Delta_{O} (v_{T})+1$.
(2) all vertices in $V - \{\hspace{0.02cm}v_S,v_T\hspace{0.02cm}\}$ satisfy Euler's condition and (3)
  if $(v_1,v_2)$ is a directed edge in $E$, then $(v_2,v_1) \not\in E$.\vspace{0.02cm}
We refer to $v_{S}$ and $v_{T}$ as the {\rm starting} and {\rm ending} vertices of
$G$, respectively.

We call $G$ a {\rm PPAD graph}
   if in addition $\Delta_{I} (v),\Delta_{O} (v) \leq 1$, for all $\ v\in V$.
\end{defi}

Edges of a PPAD graph
    form a collection of disjoint directed cycles
    and a directed path from $v_S$ to $v_T$.
In this paper, we are interested in a special family of
  PPAD graphs over $\mathbb{Z}^d_n$.
A directed graph $G=(\mathbb{Z}_n^d,E)$ is a {\rm
   generalized grid PPAD-graph  over $\mathbb{Z}_n^d$} if it is a
   generalized PPAD graph and the underlying undirected
   graph of $G$ is a subgraph of the grid graph
   defined over $\Zdn$.
Moreover, if  $G$ is also a PPAD graph, then we say $G$ is a {\rm grid PPAD graph}.

We now define the query model $\BBO_G$ for accessing a
  grid PPAD graph $G$.

\begin{defi}[\mbox{\rm $\BBO_G$}]
$\BBO_G$ is a map  from $\Zdn$ to $(\{\text{``no''}\}
\cup \bbE^d)\times (\{\text{``no''}\}
\cup \bbE^d)$ such that, for all $\vv\in \Zdn$,
\vspace{-0.05in}
\begin{itemize}
\item  $\BBO_G (\vv ) = (\text{``no''}, \vv_1-\vv)$
if $\vv$ is the starting vertex of $G$ and $(\vv,\vv_1) \in E$;
\vspace{-0.05in}
\item $\BBO_G (\vv ) = (\vv-\vv_1,\text{``no''})$
  if $\vv$ is the ending vertex of $G$ and $(\vv_1,\vv) \in E$;
\vspace{-0.05in}
\item $\BBO_G (\vv ) = (\vv-\vv_1,\vv_2-\vv)$
    if $(\vv_1,\vv)$ and $(\vv,\vv_2)$ are directed
  edges of $G$.
\vspace{-0.05in}
\item $\BBO_G(\vv)=(\text{``no''},\text{``no''})$, otherwise.
\end{itemize}
\end{defi}

In other words, $\BBO_G$ specifies the predecessor and successor
  of each vertex $\vv$ in $G$.\vspace{0.015cm}
We will use the property that if
$\BBO_G(\vv) = (\ss_1,\ss_2)$ and $\ss_1,\ss_2\in \bbE^d$,
   then $\ss_{1}+\ss_{2}\not= 0$.


Let $\text{\sf GP}^d$ be the search problem:
{\em Given a triple $(G,0^n,\uu)$,
  where $G$ is a grid PPAD graph over $\Zdn$ accessible
  by $\BBO_G$ and $\uu$ is the starting vertex of graph $G$ satisfying $u_d=1$,
  find its ending vertex.}
We use $\text{\sf RQ}_{\text{\sf GP}}^d(n)$ to denote the
  randomized query complexity for solving this problem.\vspace{-0.02cm}

\subsection{End-of-a-String}\label{}

Suppose $\Sigma$ is a finite set.
A string $S$ over $\Sigma$ of length $m$ is a sequence
  $S=a_1a_2... a_{m-1}a_m$ with $a_i\in \Sigma$.
We use $|S|=m$ to denote the length of $S$.

\begin{defi}[\mbox{\rm Non-Repeating-Strings}]
A string $S=a_1a_2...a_m$ over
  $\mathbb{Z}_n=[1:n]$ is {\rm $d$-non-repeating} for $d\in[1: m]$,
  if (1) each string over $\mathbb{Z}_n$ of length $d$
  appears in $S$ at most once; (2) $a_i$ is odd if
  $i$ is a multiple of $d$
  and $a_i$   is even  otherwise; and (3)
  $m$ is a multiple of $d$.
  We define $\eend{d}{S} = a_{m-d+1}...a_{m}$.
\end{defi}

Each $d$-non-repeating string $S=a_1...a_m$ over $\mathbb{Z}_n$
  defines a query oracle $\BBO_S$ from $\mathbb{Z}_n^d$ to
  $(\{\text{``no''}\}\cup \mathbb{Z}_n)\times (\{\text{``no''}\}\cup \mathbb{Z}_n)$: For
 $S'=b_1b_2...b_d\in \mathbb{Z}_n^d$, if $S'$ is not a substring of $S$, then $\BBO_S(S')=(\text{``no''},
 \text{``no''})$;
otherwise, there is a unique
   $k$ such that $a_{k+i-1}=b_i$, $\forall\ i\in[1:d]$. Then
$\BBO_S(S')=(\text{``no''},a_{d+1})$ if $k=1$,
$\BBO_S(S')=(a_{m-d},\text{``no''})$ if $k=m-d+1$, i.e.,
$S'=\eend{d}{S}$, and $\BBO_S(S')=(a_{k-1},a_{k+d})$, otherwise.\vspace{0.02cm}

Let $\text{\sf ES}^d$ be the search problem:
  \emph{Given a $d$-non-repeating string $S$ over $\mathbb{Z}_n$
  accessible by $\BBO_S$,
  and its first $d$ symbols $a_1a_2...a_d$ where $a_d=1$,
  find $\eend{d}{S}$.}
We let $\text{\sf RQ}_{\text{\sf ES}}^d(n)$
  denote its randomized query complexity.
It is easy to show that $\text{\sf RQ}_{\text{\sf ES}}^1(n)=\Theta(n)$.
  In section~\ref{sec:ESLOWER}, we will  prove

\begin{theo}[Complexity of \mbox{\rm $\text{\sf ES}^d$}]\label{thm:EoS}
For all sufficiently large $n$,
$$\text{\sf RQ}_{\text{\sf ES}}^d(4n+4)\hspace{0.06cm}\ge \frac{1}{2}\hspace{0.06cm}
\hspace{0.06cm}\left(\hspace{0.03cm}
\frac{n}{2\cdot 24^{d}}\hspace{0.03cm}\right)^d.\vspace{0.15cm}$$
\end{theo}

\section{Reduction Among Search Problems}\label{sec:}

In this section, we reduce $\text{\sf ES}^{d-1}$ to $\text{\sf ZP}^d$
  by first reducing $\text{\sf ES}^{d-1}$ to $\text{\sf GP}^d$
(Theorem \ref{thm:EoS2GPPAD} below) and then
  reducing $\text{\sf GP}^{d}$ to $\text{\sf ZP}^d$ (Theorem \ref{thm:GP2ZP}).
Theorem \ref{thm:FixedPoint} then follows from Theorem \ref{thm:EoS}.

\begin{theo}[\mbox{\rm From $\text{\sf ES}^{d-1}$ to $\text{\sf GP}^d$}]\label{thm:EoS2GPPAD}
For all $d\ge 2$, $\text{\sf RQ}_{\text{\sf ES}}^{d-1}(n)\le
   4d\cdot \text{\sf RQ}_{\text{\sf GP}}^d(8n+1)$.
\end{theo}

\begin{theo}[\mbox{\rm From $\text{\sf GP}^{d}$ to $\text{\sf ZP}^d$}]
   \label{thm:GP2ZP}
For all $d\ge 1$,  $\text{\sf RQ}_{\text{\sf GP}}^{d}(n)\le
    \text{\sf RQ}_{\text{\sf ZP}}^d(24n+7)$.
\end{theo}

\subsection{\mbox{\rm From $\text{\sf ES}^{d-1}$ to $\text{\sf
GP}^d$}: Proof of Theorem \ref{thm:EoS2GPPAD}}

\begin{proof} \mbox{[of Theorem \ref{thm:EoS2GPPAD}]}:
We define a map  $\calF_d$ from $\mathbb{Z}^{d-1}$
  to $\mathbb{Z}^d$:
 for $d=2$, $\calF_2(a)=(a,a)$; and for $d>2$,
  $\calF_d(\aa)=(a_1, a_1+a_2,...,a_{d-2}+a_{d-1}, a_{d-1})$.
We will crucially use the following nice property of $\calF_{d}$.
\begin{quote}
For any $k\in [1:d]$ and for any $\aa \in \mathbb{Z}^{d-1}$, we can
  uniquely determine the first $k$ and \\the last $k$ entries of $\aa $,
respectively, from the first $k$ and the last $k$ entries of $\calF_{d} (\aa )$.
\end{quote}
Let $S$ be a $(d-1)$-non-repeating string over $\mathbb{Z}_n$ of
  length $m(d-1)$ for some $m\ge 2$,  whose $(d-1)^{st}$ symbol is $1$.
We  view $S$ as a sequence of $m$ points $\aa_1,\aa_2,...\aa_m$
    in $\mathbb{Z}_n^{d-1}$, where $\aa_{i} = a_{i,1}...a_{i,d-1}$,
    such that,
$S=a_{1,1}a_{1,2}...a_{1,d-1}...a_{m,1}a_{m,2}...a_{m,d-1}.$
From $S$, we will construct a grid PPAD graph $G'$ in two stages.
In the first stage, we construct a generalized grid PPAD graph $G^*$
  over $\mathbb{Z}_{2n}^d$
  such that
\begin{description}
\item[\ \ (\textbf{A.1})] Its starting vertex is $\uu^*=\calF_d(\aa_1)$
and its ending vertex is $\ww^*=\calF_d(\aa_m)$;\vspace{-0.05cm}
\item[\ \ (\textbf{A.2})] \parbox[t]{14cm}{For every directed edge $(\uu,\vv)$
  with $\uu-\vv\in \bbE^d$,
  at most one query to $\BBO_S$ is needed to determine whether
  $(\uu,\vv)\in G^*$.}
\end{description}
Recall that a directed path is {\em simple} if it contains
  each vertex at most once.
Suppose $\uu,\vv\in \mathbb{Z}_{2n}^d$ are two vertices
that differ in only one coordinate, say the $i^{th}$ coordinate.
Suppose $\ee=(\vv-\uu)/|v_i-u_i|\in \bbE^d$.
Let $E(\uu,\vv) = \left\{
  (\uu,\uu+\ee), (\uu+\ee,\uu+2\ee),...,(\vv-\ee,\vv)\right\}$.
For $n,m_1,m_2\in \mathbb{Z}$ and $s\in \{\hspace{0.04cm}\pm 1\hspace{0.04cm}\}$,
  $(n,s)$ is {\em consistent} with $(m_1,m_2)$ if either
$m_1\le n<m_2$ and $s=+1$ or $m_2<n\le m_1$ and $s=-1$.\vspace{0.015cm}

We consider two consecutive points $\aa =\aa_{t}$ and $\bb=\aa_{t+1}$ in the
  $(d-1)$-non-repeating string $S$.
We know $\aa\not=\bb$.
We map them to vertices $\uu=\calF_d(\aa)$ and
  $\ww=\calF_d(\bb)$ in $\mathbb{Z}_{2n}^d$ and connect them
  with a path through a sequence of $(d-1)$ vertices
   $\vv_0 =\uu ,\vv_1,...,\vv_{d-1},\vv_d=\ww $
  where $v_{i,j}=u_j$ if $i<j$ and $v_{i,j}=w_j$ if $i\ge j$.
Note that $\vv_{i-1}$ and $\vv_i$ differ only in the
  $i^{th}$ coordinate.
Let $P(\aa,\bb)=\cup_{i=0}^{d-1}E\left(\vv_i,\vv_{i+1}\right)$.
Then $P(\aa,\bb)$ is a simple directed
  path in the grid graph over $\mathbb{Z}_{2n}^d$
  from $\uu=\vv_0$ to $\ww=\vv_d$.
As $S$ is $(d-1)$-non-repeating, $\aa_1\not=\aa_m$.
By Property \ref{lem:gppad}, Proposition \ref{lem:cases} and Lemma \ref{lem:structural} below,
$G^*=(\mathbb{Z}^d_{2n},\cup_{i=1}^{m-1} P(\aa_i,\aa_{i+1}))$
  is a generalized grid PPAD graph. See Figure \ref{fig2d} for an
  example.

\begin{figure}[!t]
\begin{center}
\includegraphics[height=3.8cm]{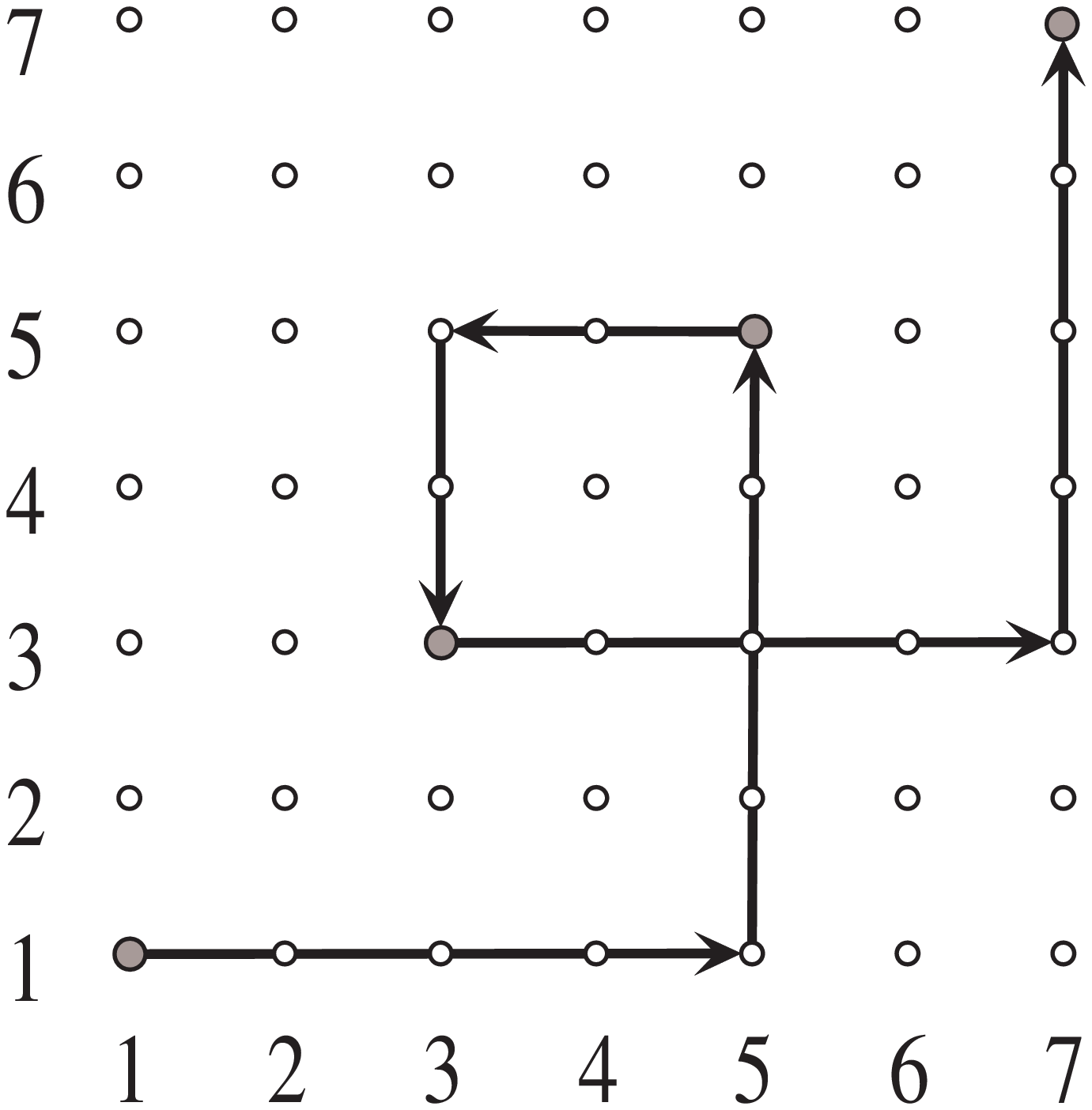}\hspace{1.0cm}
\includegraphics[height=3.8cm]{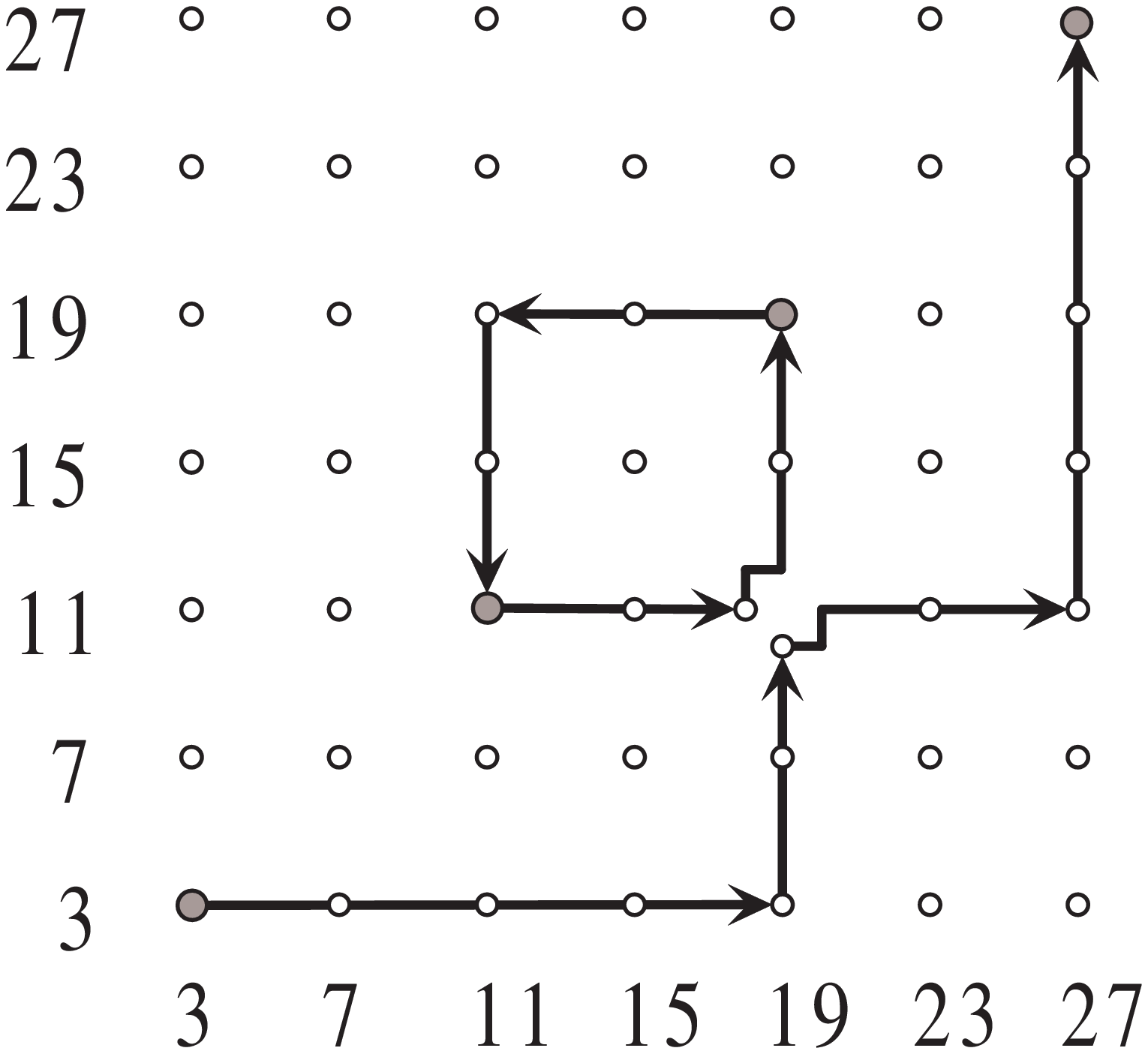}\vspace{-0.15cm}
\end{center}
\vspace{-0.05in}
\caption{Graph $G^*$ and $G'$ constructed from string $1537$}\vspace{-0.06in}\label{fig2d}
\end{figure}

\begin{prop}[\mbox{\rm Path Union}]\label{lem:gppad}
Let $P_1,P_2,...,P_m$ be $m$ simple directed paths over $V$
such that
(1) each path has length at least one, (2) the ending vertex of
  $P_i$ is same as the starting vertex of $P_{i+1}$,
 (3) the starting vertex of $P_1$ is different from the ending vertex of $P_m$,
  and (4) if $(u,v)\in P_i$, then $(u, v),(v,u)\notin P_j$, $\forall j\not= i$.
Then $G=(V,\cup_{i=1}^m P_i)$ is a generalized PPAD~graph.
\end{prop}

\begin{prop}[\mbox{\rm Local Characterization of $P(\aa,\bb)$}]
  \label{lem:cases}
For $\vv\in \mathbb{Z}_{2n}^d$ and $s\in \{\pm 1\}$,
\begin{enumerate}
\vspace{-0.05in}
\item  $(\vv,\vv+ s\ee_1)\in P(\aa,\bb)$ if and only if
  $(v_1,s)$ is consistent with $(a_1,b_1)$, $a_{d-1}=v_d$,
  and $a_{d-i}=v_{d-i+1}-a_{d-i+1}$ for all $2\le i\le d-1$;
\vspace{-0.05in}
\item $(\vv,\vv+s\ee_d)\in P(\aa,\bb)$ if and only if
  $(v_d,s)$ is consistent with $(a_{d-1},b_{d-1})$, $b_1=v_1$,
   and $b_i=v_i-b_{i-1}$, for $2\le i\le d-1$; and
\vspace{-0.05in}
\item for $1<k<d$, $(\vv,\vv+s\ee_k)\in P(\aa,\bb)$
   if and only if $(v_k,s)$
   is consistent with $(a_{k-1}+a_k,b_{k-1}+b_k)$ and \emph{(3.1)}
     $a_{d-1}=v_d$, and $a_{d-i}=v_{d-i+1}-a_{d-i+1}$ for $2\le i\le d-k$ and
     \emph{(3.2)} $b_1=v_1$, and $b_i=v_i-b_{i-1}$ for $2\le i\le k-1$.
\vspace{-0.05in}
\end{enumerate}
\end{prop}

\begin{lemm}[\mbox{\rm Structural Correctness}]\label{lem:structural}
For all $(d-1)$-non-repeating string $S=\aa_1\aa_2...\aa_m$
  over $\mathbb{Z}_n$, if $(\uu,\vv)\in P(\aa_i,\aa_{i+1})$ then
   $(\uu,\vv),(\vv,\uu)\notin P(\aa_j,\aa_{j+1})$ for all $i\neq j$.
\end{lemm}

\begin{proof}
We only prove the case when
  $\ee=\vv-\uu=s\ee_k$ with $1<k<d$
  and $s\in \{\pm 1\}$.
The other two cases are similar.
From Proposition~\ref{lem:cases},
   $(\uu,\vv)\in P(\aa_i,\aa_{i+1})$ implies that $\aa_i$ and $\aa_{i+1}$ satisfy
  conditions (3.1) and (3.2).
If $(\uu,\vv)$ or $(\vv,\uu)$ is in $P(\aa_{j},\aa_{j+1})$, then $\aa_j$ and
  $\aa_{j+1}$ also satisfy these two conditions.
Then
$a_{i,k}a_{i,k+1}...a_{i,d-1}a_{i+1,1}...a_{i+1,k-1}=a_{j,k}a_{j,k+1}...
  a_{j,d-1}a_{j+1,1}...a_{j+1,k-1},$ which
  contradicts with the assumption that $S$ is $(d-1)$-non-repeating.
\end{proof}

We prove Property \textbf{A.2} as follows.

\begin{proof}[Proof of Property \rm\textbf{A.2}]
We will only prove for the case when
   $\ee=\vv-\uu=s\ee_k$ with $1<k$ $<d$.
The other two cases are similar and simpler.
To determine whether $(\uu,\vv)\in G^*$ or not,
   we consider the string $S'=a_ka_{k+1}...a_{d-1}b_1...b_{k-1}$
  that satisfies both (3.1) and (3.2) in Proposition~\ref{lem:cases}.
  Edge $(\uu,\vv)\in G^*$ if and only if 1)
  $S'\in \mathbb{Z}^{d-1}_n$; 2) $a_{d-1}$ is odd; 3)
   $\BBO_S(S')=(a,b)$ for some $a,b\in \mathbb{Z}_n$; and 4)
  $(u_k,s)$ is consistent with $(a+a_k,b_{k-1}+b)$,
So, only one query to $\BBO_S$ is needed.
\end{proof}

 In the second stage, we construct a grid PPAD graph $G'$
  over $\mathbb{Z}^d_{8n+1}$ from graph $G^*$.
Let $\Gamma(\vv)=4\vv-\mathbf{1}$ for all $\vv \in \mathbb{Z}_{2n}^d$.
Our $G'$ will satisfy the following two properties.
See Figure \ref{fig2d} for an example.
\begin{description}
\item[\ \ (\textbf{B.1})] Its starting vertex is
   $\uu'=\Gamma(\uu^*)-2\ee_d$; its ending vertex $\ww'$
   satisfies $\|\ww'-\Gamma(\ww^*)\|\le 1$;\vspace{-0.05cm}
\item[\ \ (\textbf{B.2})] \parbox[t]{14.3cm}{For each $\vv\in \mathbb{Z}^d_{8n+1}$,
  one can determine $\BBO_{G'}(\vv)$ from the predecessors and successors
  of $\uu$ in $G^*$, where
  $\uu$ is the lexicographically smallest vertex such that $\| \vv-\Gamma(\uu)\le 2 \|$.}
\end{description}


Two subsets $H_1$ and $H_2$ of $\bbE^d$, where $d\ge 2$,
   form a {\em balanced-non-canceling pair}
   if $|H_1|=|H_2|$ and $\ss_1+\ss_2\not=\00$ for all $\ss_1\in H_1$
  and $\ss_2\in H_2$.
Let $H_I(\uu) = \{\hspace{0.04cm}\ee \in \Ed\ |\ (\uu-\ee,\uu) \in E^* \hspace{0.04cm}\}$
  be the vector differences of $\uu$ and its predecessors in $G^{*}$.
Similarly, let $H_O(\uu) = \{\hspace{0.04cm}\ee \in \Ed\ |\ (\uu,\uu+\ee) \in E^*\hspace{0.04cm} \}$
  be the vector differences of the successors of $\uu$ and $\uu$.
In the construction below, we will use the fact that
  if $\uu$ satisfies Euler's condition then $(H_I,H_O)$ is
  a balanced-non-canceling pair.\vspace{0.02cm}

Using the procedure of Figure~\ref{graph},
  we build a  graph $G[H_1,H_2]=( \{\hspace{0.02cm}-1,
  0,+1\hspace{0.02cm}\}^d,E [H_1,H_2] )$
  for each balanced-non-canceling pair $H_1$ and $H_2$.
$G[H_1,H_2]$ has the following properties:
(1)  For every $\uu\in \setof{-1,0,+1}^d$,
   $\Delta_I(\uu),\Delta_O(\uu)\le 1$;
(2) A vector $\uu\in \setof{-1,0,+1}^d$ has $\Delta_I(\uu)=0$ and
  $\Delta_O(\uu)=1$ iff there exists an $\ee\in H_1$ such that
   $\uu=\00-\ee$; (3)
   A vector $\uu\in \setof{-1,0,+1}^d$ has $\Delta_I(\uu)=1$ and
  $\Delta_O(\uu)=0$ iff there exists an $\ee\in H_2$ such that
    $\uu=\00+\ee$.\vspace{0.02cm}

\begin{figure}[t]
\rule{\textwidth}{1pt}\vspace{0.05cm}

\textbf{\ \ Graph $G[H_1,H_2]$,
  where $(H_1,H_2)$ is a balanced-non-canceling pair}

\vspace{-0.2cm}\rule{\textwidth}{1pt}\vspace{0.16cm}

\begin{tabular}{@{\hspace{0.1cm}}r@{\hspace{0.2cm}}p{\textwidth}}
\ \ 1\hspace{0.05cm}: & set edge set $E[H_1,H_2]=\emptyset$ \\ [0.5ex]

\ \ 2\hspace{0.05cm}: & \textbf{while} $H_1\not=\emptyset$ \textbf{do} \\ [0.5ex]

\ \ 3\hspace{0.05cm}: & \ \ \ \ let $\ss_1$ be the smallest vector
  in $H_1$ and $\ss_2$ be the largest vector in
$H_2$\\ [0.5ex]

& \ \ \ \ according to the lexicographical ordering;\\[0.5ex]

\ \ 4\hspace{0.05cm}: & \ \ \ \ set $H_1=H_1-\{\hspace{0.03cm}\ss_1\hspace{0.03cm}\}$ and
$H_2=H_2-\{\hspace{0.03cm}\ss_2\hspace{0.03cm}\}$; \\[0.5ex]

\ \ 5\hspace{0.05cm}: & \ \ \ \ set $E[H_1,H_2]=E[H_1,H_2]
   \cup\{\hspace{0.03cm}(\00-\ss_1,\00-\ss_1+\ss_2),(\00-\ss_1+\ss_2,
\00+\ss_2)\hspace{0.03cm}\}$\\[0.29ex]
\end{tabular}

\vspace{0.097cm}\rule{\textwidth}{1pt} \caption{Construction of
Graph $G[H_1,H_2]=(\{-1,0,+1\}^d, E[H_1,H_2])$} \label{graph}
\end{figure}

Let $\uu^*$ be the starting vertex and $\ww^*$ be the
  ending vertex of $G^*$.
We build a grid PPAD graph $G'=(\mathbb{Z}_{8n+1}^d,E')$
  by applying~the procedure of Fig.~\ref{graph} locally
  to every vertex $\uu \in \mathbb{Z}_{2n}^d$ of $G^*$.
We use $(H_I(\uu),H_O(\uu))$ or a slight modification of $(H_I(\uu),H_O(\uu))$
  when $\uu=\uu^*$ or $\ww^*$.
  Initially we set $E' = \emptyset$. Recall $\Gamma(\uu)=4\uu -\11$.

\begin{enumerate}
\item {\bf [\hspace{0.06cm}local embedding of the starting vertex\hspace{0.04cm}]}
Since $u^*_d=1$, we have $\ee_d\notin H_I(\uu^*)$ and $-\ee_d\notin H_O(\uu^*)$.
Let $H_I=H_I(\uu^*)\cup\{\ee_d\}$.
We add edges $(\Gamma(\uu^*)-2\ee_d,\Gamma(\uu^*)-\ee_d)$
  and $(\Gamma(\uu^*)+\ss_1,\Gamma(\uu^*)+\ss_2)$ to $E'$
  for all edges $(\ss_1,\ss_2)$ in $G[H_I,H_O(\uu^*)]$.\vspace{-0.04cm}

\item
 {\bf [\hspace{0.06cm}local embedding  of the ending vertex\hspace{0.04cm}]}
As $|H_I(\ww^*)|=|H_O(\ww^*)|+1$, $H_I(\ww^*)\not= \emptyset$.
Let $\ee $ be the smallest vector in  $H_I(\ww^*)$,
    and $H_I=H_I(\ww^*)-\{\hspace{0.03cm}\ee\hspace{0.03cm} \}$.
Add edges $(\Gamma(\ww^*)+\ss_1,\Gamma(\ww^*)+\ss_2)$ to $E'$ for all edges
   $(\ss_1,\ss_2)$ in $G[H_I,H_O(\ww^*)]$.\vspace{-0.04cm}

\item
 {\bf [\hspace{0.06cm}local embedding of other vertices\hspace{0.04cm}]}
For each  $\uu\in G^*$, add $(\Gamma(\uu)+\ss_1,\Gamma(\uu)+\ss_2)$
  to $E'$ for all edges $(\ss_1,\ss_2)$ in $G[H_I(\uu),H_O(\uu)]$.\vspace{-0.04cm}

\item
 {\bf [\hspace{0.06cm}connecting local embeddings\hspace{0.04cm}]}
For each edge $(\uu,\vv)\in G^*$,
  let $\ee=\vv-\uu\in \bbE^d$.
We add $(\Gamma(\uu)+\ee,\Gamma(\uu)+2\ee)$ and $(\Gamma(\uu)+2\ee,\Gamma(\uu)+3\ee)$
   to $E'$.
\end{enumerate}

\noindent It is quite mechanical to check that $G'$ is a PPAD grid graph that satisfies
   both Property \textbf{B.1} and \textbf{B.2}.
We therefore complete the proof of Theorem \ref{thm:EoS2GPPAD}.
\end{proof}



\subsection{Canonicalization of Grid-PPAD Graphs}\label{}

To ease our reduction from a grid-PPAD graph to
  a Brouwer function,
  we first canonicalize the grid-PPAD graph by
  regulating the way its path starts, moves, and ends.

\begin{defi}[\mbox{\rm Canonical Grid-PPAD Graphs}]
A grid-PPAD graph $G$ over $\Zdn$ for $d\ge 2$ and $n>1$ is {\rm canonical}
if $\BBO_G$ satisfies $\BBO_G(\uu) \in \calS^d$ for all $\uu \in \Zdn$,
  where
\begin{equation*}
\calS^2=\{(\text{``no''},\text{``no''}),(\text{``no''},\ee_{2}),(\ee_{2},\text{``no''})\}
\cup\big\{\hspace{0.05cm}(\ss_{1},\ss_{2})\hspace{0.1cm}\big|\hspace{0.1cm}\ss_{1},\ss_{2}\in
\bbE^2,\ss_{1}+\ss_{2}\not=0\hspace{0.05cm}\big\}.
\end{equation*}
For $d\ge 3$, $\calS^d$ is the smallest subset of $(\{\text``no''\}\cup\bbE^d)\times
(\{\text``no''\}\cup\bbE^d)$ that satisfies:
\begin{enumerate}
\item $(\text{``no''},\text{``no''}),(\text{``no''},\ee_{d}),
   (\ee_{d},\text{``no''})\in \calS^d$;\vspace{-0.08cm}
\item $\{\ee_{d}\}\times \{\ee_{d-1},\ee_{d}\}\subset \calS^d$ and $\{-\ee_{d}\}
  \times \{\ee_{d-1},-\ee_{d}\}\subset \calS^d$;\vspace{-0.08cm}
\item  $\{\ee_{k}\}\times \{\ee_{k-1},\ee_{k},\pm
   \ee_{k+1}\}\subset \calS^d$ and $\{-\ee_{k}\}\times\{\ee_{k-1},-\ee_{k}\}\subset \calS^d$,
  for $3\le k< d$;\vspace{-0.08cm}
\item $\{\ee_{2}\}\times \{\pm \ee_{1},\ee_{2},\pm \ee_{3}\}\subset
   \calS^d$ and $\{-\ee_{2}\}\times \{\pm \ee_{1},-\ee_{2}\}\subset \calS^d$;\vspace{-0.08cm}
\item $\{\ee_{1}\}\times \{\ee_{1},\pm\ee_{2}\}\subset \calS^d$; and
   $\{-\ee_{1}\}\times \{-\ee_{1},\pm \ee_{2}\}\subset \calS^d$.
\end{enumerate}
\end{defi}

\begin{figure}[!h]
\begin{center}\vspace{-0.08cm}
\includegraphics[height=4cm]{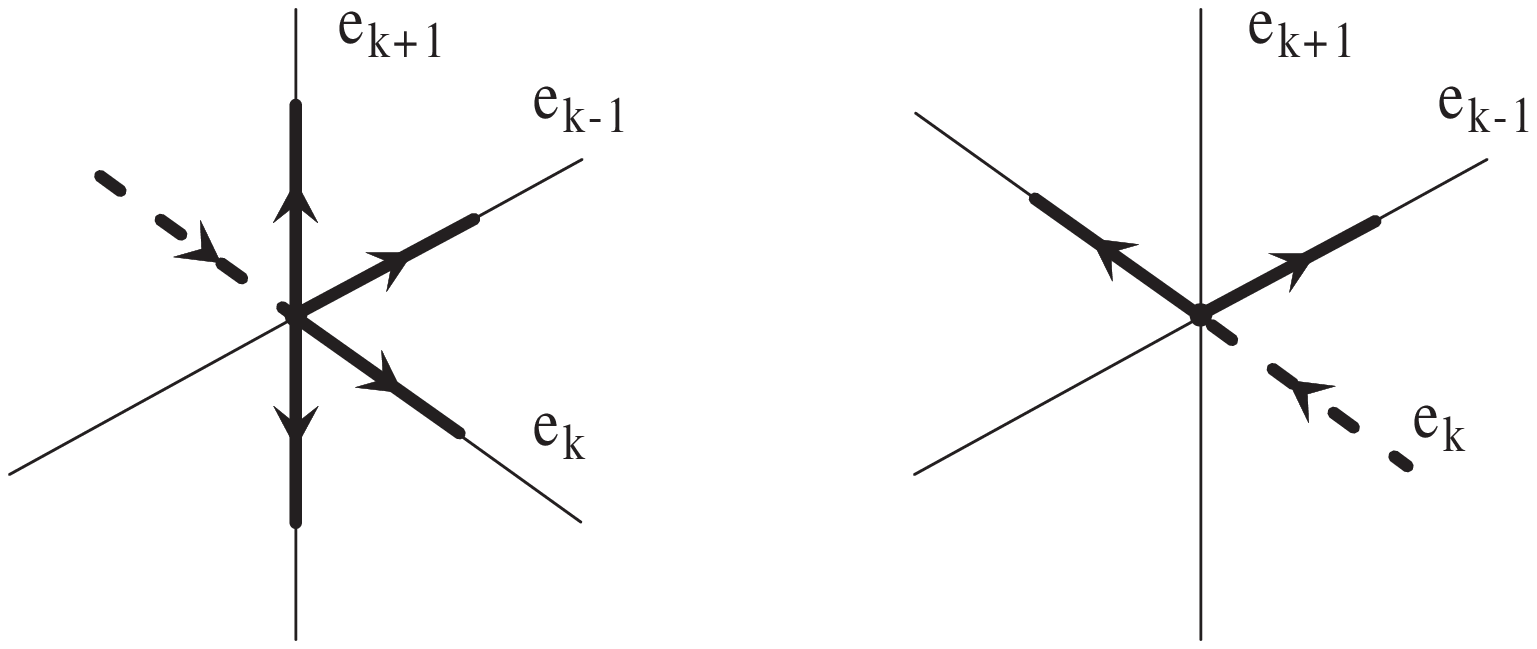}\vspace{-0.6cm}
\end{center}
\caption{ }\label{fig2da}\vspace{-0.2cm}
\end{figure}

Informally, edges in a canonical grid-PPAD graph
   over $\Zdn$ contains a single directed path
   starting at a point $\uu \in \Zdn$ with $u_{d} = 1$ and
   ending at a point, say $\ww$, and possibly some cycles.
The second vertex on the path is $\uu +\ee_{d}$
    and the second-to-the-last vertex is $\ww-\ee_d$.
The path and the cycles satisfy
   the following conditions (below we will abuse ``path'' for both
``path'' and ``cycle''):
(1) To follow a directed edge along $\ee_k$ (for $k\ge 3$),
  the path can only move
    locally in a 3D framework defined by
    $\{ \ee_{k-1},\ee_k,\pm \ee_{k+1}\}$, see Figure \ref{fig2da}, (for $k=d$,
    it can only move in a 2D framework).
In a way, we view the $d$-dimensional space as a nested ``affine subspaces'' defined
  by $\{\pm \ee_1,...,\pm \ee_k\}$ for $1\leq k \leq d$.
So to follow a positive principle direction $\ee_k$,
  the path can {\em move down} a dimension along the positive
  direction $\ee_{k-1}$, stay continuously along $\ee_k$, or
  {\em move up} a dimension (unless $k=d$) along
  either $\pm \ee_{k+1}$.
(2) To follow a directed edge along $-\ee_k$ for $k\ge 3$,
   the path can only move
    locally in a 2D framework defined by
    $\{\ee_{k-1},-\ee_k\}$, see Figure \ref{fig2da}.
The path can {\em move down} a dimension along the positive
  direction of $\ee_{k-1}$ or stay continuously along $-\ee_k$,
  but it is not allowed to {\em move up} or {\em leave}
  this $k$-dimensional ``affine subspace''.
In the $\{\pm \ee_1,\pm\ee_2 \}$ framework, the path is less restrictive as defined
   by conditions 4 and 5.

In other words, the path can not move-up from an ``affine
  subspace'' (with the exception of the $\pm\ee_1$ space)
  without first taking a step along the highest positive principle direction in
  the subspace.
Similarly, the path can only move-down to an ``affine subspace''
  by taking a positive first step along its highest principle direction.
Otherwise, the path moves continuously.
\vspace{0.02cm}

Let $\text{\sf CGP}^d$ be:\hspace{-0.08cm}
{\em Given a triple $(G,0^n,\uu^*)$
  where $G$ is a canonical grid-PPAD graph over $\Zdn$ accessible
  by $\BBO_G$ and $\uu^*$ is the starting vertex of $G$ with $u^*_d=1$,
  find the ending vertex of $G$.}
We use $\text{\sf RQ}_{\text{\sf CGP}}^d(n)$ to denote the
  randomized query complexity for solving this problem.\vspace{0.01cm}

We now reduce  $\text{\sf GP}^d$  to  $\text{\sf CGP}^d$.
Before stating and proving this result, we first
  give two geometric lemmas.
They provide the local
  operations for canonicalization.\vspace{0.01cm}

We start with some notation.
A sequence $P=\uu_1...\uu_m$, for $m\ge 2$, is a {\em canonical local
  path} if $\uu_i$'s are distinct elements from
 $\{-3,-2,...,+2,+3\}^d$ and $(\uu_i-\uu_{i-1},\uu_{i+1}-
\uu_i)\in \calS^d$ for all $2\le i\le m-1$.
Suppose $P=\uu_1\uu_2...\uu_m$ and $Q=\vv_1\vv_2...\vv_k$ are two paths
  with $\uu_m=\vv_1$.
We use $P\diamond Q$ to denote their ``concatenation'':
    $P\diamond Q=\uu_1\uu_2...\uu_m\vv_2...\vv_k$.

\begin{lemm}[\mbox{\rm Ending Gracefully}]\label{lem:ending}
For each $\ss\in \bbE^d$ with $d\ge 2$,
  there is a canonical local path
  $P[d,\ss]$ $=\uu_1... \uu_m$ satisfying
  $\uu_m-\uu_{m-1}=\ee_d$, $\uu_1=-3\ss$, $\uu_2=-2\ss$, and
    $\forall\ i\in[2:m]$,~$\|\uu_i\|\le 2$.
\end{lemm}
\begin{proof}
We consider the three cases: $\ss=\ee_l$ for $1\le l\le d$,
  $\ss=-\ee_1$ and $\ss=-\ee_l$ for $2\le l\le d$.\vspace{0.0cm}

In the first case, we set $P[d,\ss]=\uu_1\uu_2...\uu_{d-l+2}$ where
  $\uu_1=-3\ee_l$,
  $\uu_2=-2\ee_l$ and $\uu_{i}=\uu_{i-1}+\ee_{l+i-2}$ for $3\le i\le d-l+2$.
In the second case, we set  $P[d,\ss]=\uu_1\uu_2...\uu_d\uu_{d+1}$ where
  $\uu_1=3\ee_1$, $\uu_2=2\ee_1$ and
  $\uu_i=\uu_{i-1}+\ee_{i-1}$ for $3\le i\le d+1$.
In the third case, we set $P[d,\ss]=\uu_1\uu_2...\uu_{d-l+5}$
  where $\uu_1=3\ee_l$, $\uu_2=2\ee_l$, $\uu_3=\ee_l$ and
   $\uu_i=\uu_{i-1}+\ee_{l+i-5}$ for $4\le i\le d-l+5$.\vspace{0.005cm}

One can easily check that $P[d,\ss]$ satisfies the conditions of the lemma.
\end{proof}

\begin{lemm}[\mbox{\rm Moving Gracefully}]\label{lem:walking}
For all $\ss_1,\ss_2\in \bbE^d$ with $d\ge 2$ such that
  $\ss_1+\ss_2\not=0$, there exists a canonical local path
$P[d,\ss_1,\ss_2]=\uu_1\uu_2...\uu_{m-1}\uu_m$ that satisfies
  $\uu_1=-3\ss_1$, $\uu_2=-2\ss_1$, $\uu_{m-1}= 2\ss_2$,
  $\uu_m=3\ss_2$, and $\forall\ i\in [2: m-1]$, $\|\uu_i\|\le 2$.
\end{lemm}
\begin{proof}
We prove by induction on $d$ that there is a canonical local path
$P[d,\ss_1,\ss_2]$ such that\vspace{-0.01cm}
\begin{enumerate}
\item $P[d,\ss_1,\ss_2]$ satisfies all the conditions in the statement
  of the lemma;\vspace{-0.06cm}

\item For all $\ss_1,\ss_2\in\bbE^d$ such that $\ss_1+\ss_2\not=0$,
$\ee_d,-\ee_d \notin P[d,\ss_1,\ss_2]$; and\vspace{-0.06cm}

\item If $\ss_1=-\ee_d$, then the first $3$ vertices of
  $P[d,-\ee_d,\ss_2]$ are $3\ee_d$, $2\ee_d$ and $(\ee_{d-1}+2\ee_d)$.
If $\ss_2=-\ee_d$, then the last $3$ vertices of
  $P[d,\ss_1,-\ee_d]$ are $(-\ee_{d-1}-2\ee_d)$, $-2\ee_d$ and $-3\ee_d$.\vspace{-0.01cm}
\end{enumerate}
The base case when $d=2$ is trivial.
Inductively we assume, for $2\le d'< d$,
  path $P[d',\ss'_1,\ss'_2]$ exists for all $\ss'_1,\ss'_2\in \bbE^{d'}$
  such that $\ss'_1+\ss'_2\not=0$.\vspace{0.02cm}
We let $P'[d',\ss'_1,\ss'_2]$ denote
  the sub-path of $P[d',\ss'_1,\ss'_2]$ such that
$$P[d',\ss'_1,\ss'_2]=(-3\ss'_1)\hspace{0.04cm}(-2\ss'_1)\hspace{0.04cm} \diamond\hspace{0.04cm} P'[d',\ss'_1,\ss'_2]
  \hspace{0.04cm}\diamond\hspace{0.04cm} (2\ss'_2)\hspace{0.04cm} (3\ss'_2).$$
Note that $P'[d',\ss'_1,\ss'_2]$ starts
  with $-2\ss'_1$ and ends with $+2\ss'_2$.\vspace{0.015cm}

We will use $D$ to denote the map  $D(\rr)=(r_1, ...,r_{d-1})$
 from $\mathbb{Z}^{d}$ to $\mathbb{Z}^{d-1}$
 and $U$ to denote the map $U(\rr)=(r_1, ...,r_{d-1},0)$
 from $\mathbb{Z}^{d-1}$ to $\mathbb{Z}^{d}$.
For a canonical local path $P=\uu_1 ... \uu_m$ in
  $\{-3,...,+3\}^{d-1}$, we use $U(P)$ to denote path
  $U(\uu_1)U(\uu_2)...U(\uu_m)$ in $\{-3,...,+3\}^{d}$.\vspace{0.015cm}

For $\ss_1,\ss_2\in \bbE^{d}$
  with $\ss_1+\ss_2\not=0$,  we use the following procedure to build
  $P[d,\ss_1,\ss_2]$. Let
\begin{eqnarray*}
P^+\hspace{-0.2cm} & = &\hspace{-0.1cm}(-3\ee_{d})\hspace{0.04cm}(-2\ee_{d})
\hspace{0.04cm}(\ee_{d-1}-2\ee_{d})\hspace{0.04cm}
(\ee_{d-1}-\ee_{d})\hspace{0.04cm}(\ee_{d-1})\hspace{0.04cm}(2\ee_{d-1});\\[0.3ex]
P^-\hspace{-0.2cm} & = &\hspace{-0.1cm}(3\ee_{d})\hspace{0.04cm}(2\ee_{d})\hspace{0.04cm}
(\ee_{d-1}+2\ee_{d})\hspace{0.04cm}(\ee_{d-1}+\ee_{d})\hspace{0.04cm}(\ee_{d-1})\hspace{0.04cm}(2\ee_{d-1});\\[0.3ex]
Q^+\hspace{-0.2cm} &= &\hspace{-0.1cm}(-2\ee_{d-1})\hspace{0.04cm}(-\ee_{d-1})
\hspace{0.04cm}(-\ee_{d-1}+\ee_{d})\hspace{0.04cm}(-\ee_{d-1}
  +2\ee_{d})\hspace{0.04cm}(2\ee_{d})\hspace{0.04cm}(3\ee_{d});\\[0.3ex]
Q^-\hspace{-0.2cm} & = & \hspace{-0.1cm}(-2\ee_{d-1})\hspace{0.04cm}(-\ee_{d-1})\hspace{0.04cm}
(-\ee_{d-1}-\ee_{d})\hspace{0.04cm}(-\ee_{d-1}-
  2\ee_{d})\hspace{0.04cm} (-2\ee_{d})\hspace{0.04cm}(-3\ee_{d}).
\end{eqnarray*}
\begin{itemize}
\item [1.]
  If $\ss_1\notin \{\pm \ee_{d}\}$, then set
 $P=(-3\ss_1)(-2\ss_1)$ and $\ss_1'=D(\ss_1)$;\vspace{-0.1cm}

Set $P=P^+$ and $\ss_1'=-\ee_{d-1}$ if $\ss_1=\ee_{d}$
  and $P=P^-$ and $\ss_1'=-\ee_{d-1}$ if $\ss_1=-\ee_{d}$;

\item [2.]
If $\ss_2\notin \{\pm \ee_{d}\}$, then set $Q=(2\ss_2)(3\ss_2)$ and $\ss_2'=D(\ss_2)$;\vspace{-0.1cm}

Set $Q=Q^+$ and $\ss_2'=-\ee_{d-1}$ if $\ss_2=\ee_{d}$ and
 $Q=Q^-$ and $\ss_2'=-\ee_{d-1}$ if $\ss_2=-\ee_{d}$;\vspace{-0.06cm}
\item [3.] Set $P[d,\ss_1,\ss_2]=P\hspace{0.04cm} \diamond
\hspace{0.04cm} U(P'[d-1,\ss_1',\ss_2']) \hspace{0.04cm}\diamond\hspace{0.04cm} Q$:
\end{itemize}
One can check that $P[d,\ss_1,\ss_2]$ satisfies all three
conditions of the inductive statement.
\end{proof}

\begin{theo}[\mbox{\rm Canonicalization}]\label{thm:cannon}
For all $d\ge 2$,
  $\text{\sf RQ}_{\text{\sf GP}}^d(n)\le \text{\sf RQ}_{\text{\sf CGP}}^d(6n+1)$.
\end{theo}
\begin{proof}
Let $\Gamma(\uu)=6\uu-\22$ be a map from $\mathbb{Z}^d$ to $\mathbb{Z}^d$.
Given any grid-PPAD graph $G^*$ over $\mathbb{Z}^d_n$,
  we now use the canonical local paths provided in Lemmas \ref{lem:ending} and
  \ref{lem:walking}
  to build a canonical grid-PPAD graph $G = (\mathbb{Z}_{6n+1}^d,E)$.
In the procedure below, initially $E = \emptyset$:\vspace{-0.02cm}
\begin{enumerate}
\item {\bf [\hspace{0.06cm}canonicalizing the starting vertex\hspace{0.04cm}]}:
  Let $\uu^*$ be the starting vertex of $G^*$\hspace{-0.1cm}.
  Suppose $\BBO_{G^*}(\uu^*)=(\text{``no''},\ss)$.
  As $u^*_d=1$, we have $\ss\not=-\ee_d$.
  For every edge $(\vv_1,\vv_2)$ appears in path
  $P[d,\ee_d,\ss]$, add $(\Gamma(\uu^*)+\vv_1,\Gamma(\uu^*)+\vv_2)$ to $E$;\vspace{-0.08cm}

\item {\bf [\hspace{0.06cm}canonicalizing the ending vertex\hspace{0.04cm}]}:
Let $\ww^*$ be the ending vertex of $G^*$.
Suppose $\BBO_{G^*}(\ww^*)=(\ss,\text{``no''})$.
For every  $(\vv_1,\vv_2)$ in $P[d,\ss]$, add
   $(\Gamma(\ww^*)+\vv_1,\Gamma(\ww^*)+\vv_2)$ to $E$;\vspace{-0.08cm}

\item {\bf [\hspace{0.06cm}canonicalizing other vertices\hspace{0.04cm}]}:
For all $\uu\in \mathbb{Z}^d_n-\{\uu^{*},\ww^{*}\}$,
  if $\BBO_{G^*}(\uu)=(\ss_1,\ss_2)$ and $\ss_1,\ss_2\not=\text{``no''}$, then
  add $(\Gamma(\uu)+\vv_1,\Gamma(\uu)+\vv_2)$ to $E$ for every edge
  $(\vv_1,\vv_2)$ in $P[d,\ss_1,\ss_2]$.\vspace{-0.02cm}
\end{enumerate}
By Lemmas \ref{lem:ending}, \ref{lem:walking} and the procedure above,
  $G = (\mathbb{Z}_{6n+1}^d,E)$ is a canonical grid-PPAD graph
  that satisfies the following two properties, from which Theorem
  \ref{thm:cannon}  follows.\vspace{-0.04cm}
\begin{description}
\item[\ \ (\textbf{C.1})]
\parbox[t]{13.5cm}{The starting vertex of $G$ is $\Gamma(\uu^*)-3\ee_d$,
   and the ending vertex $\ww$ of $G$ satisfies $\|\ww-\Gamma(\ww^*)\|\le 2$.}\vspace{-0.05cm}

\item[\ \ (\textbf{C.2})]
\parbox[t]{14cm}{For every vertex
  $\vv\in G$, to determine $\BBO_{G}(\vv)$, one only need
  to know $\BBO_{G^*}(\uu)$ where $\uu$ is the lexicographically smallest vertex in
  $\mathbb{Z}_n^d$ such that $\|\vv-\Gamma(\uu)\|\le 3$.}\vspace{-0.25cm}
\end{description}
\vspace{-0.25cm}
\end{proof}

\subsection{From $\text{\sf CGP}^{d}$ to $\text{\sf ZP}^d$: 
  Complete the Proof of Theorem~\ref{thm:GP2ZP}
}\label{}

Now we reduce $\text{\sf CGP}^d$ to $\text{\sf ZP}^d$.
The main task of this section is to, given a canonical grid-PPAD graph $G= (\Zdn,E)$
   and its starting vertex $\uu^{*}$,
   construct a discrete Brouwer function $f_G:
   \ZZ{d}{4n+2}\rightarrow \{\hspace{0.04cm}\00\hspace{0.04cm}\}\cup\bbE^d$ that
   satisfies the following two properties:\vspace{-0.02cm}
\begin{description}
\item[\ \ \textbf{(D.1)}] \parbox[t]{14cm}{$f_G$ has exactly one zero point $\rr^*$, and
  $\Psi^{-1}(\rr^*)$ is the ending vertex of $G$, and}\vspace{-0.05cm}

\item[\ \ (\textbf{D.2})] \parbox[t]{14cm}{For each $\rr\in \ZZ{d}{4n+2}$,
 at most one query to $\BBO_G$ is needed to
 evaluate $f_G$,}\vspace{-0.02cm}
\end{description}
where $\Psi(\uu)=4\uu$ is a
  map from $\ZZ{d}{n}$ to $\ZZ{d}{4n+2}$. Immediately from these properties, we have

\begin{theo}[\mbox{\rm From $\text{\sf CGP}^d$ to $\text{\sf ZP}^d$}]\label{thm:toBrouwer}
For~all $d\ge 2$,
$\text{\sf RQ}_{\text{\sf CGP}}^d(n) \le
   \text{\sf RQ}_{\text{\sf ZP}}^d(4n+2).$
\end{theo}

\subsubsection*{Local Geometry of Canonical Grid PPAD Graphs}

To construct a Brouwer function from $G = (\Zdn,E)$,
  we define a set $\calI_G\subset \ZZ{d}{4n+2}$,
   which looks like a collection of pipes,
   to embed and insulate the component of $G$.
$\calI_G$ has two parts, a kernel $\calK_G$ and a boundary $\calB_G$.
We will first construct a direction-preserving
   function $f^G$ on $\calB_G$.
We then extend the function onto $\ZZ{d}{4n+2}$ to define $f_G$.
In our presentation, we will
  only use $\uu,\vv,\ww$ to denote vertices in $\ZZ{d}{n}$
   and use $\pp,\qq,\rr$ to denote points in $\ZZ{d}{4n+2}$.\vspace{0.02cm}
Let $\uu$ and $\vv$ be two vertices in $\ZZ{d}{n}$
    with $\uu-\vv\in \bbE^d$.
We abuse $\Psi(\uu\vv)$ to denote the set of five integer
  points on line segment $\Psi(\uu)\Psi(\vv)$.
Let $\uu^*$ and $\ww^*$ be the starting and ending vertices of $G$.
We define
\begin{eqnarray*}
&\calK_G=\left(\hspace{0.05cm}\bigcup_{(\vv_1,\vv_2)\in E}\Psi(\vv_1\vv_2)
  \hspace{0.05cm}\right)\cup
  \Big\{\hspace{0.08cm}\Psi(\uu^*)-\ee_d,\Psi(\ww^*)+\ee_d\hspace{0.08cm}\Big\},&
\end{eqnarray*}
\begin{eqnarray*}
&\calB_G= \Big\{\hspace{0.08cm}\rr\notin \calK_G\hspace{0.1cm}\Big|
  \hspace{0.1cm}\exists\
  \rr'\in \bigcup_{(\vv_1,\vv_2)\in E}\Psi(\vv_1\vv_2), \|\rr- \rr'\| =1
\hspace{0.08cm}\Big\},&
\end{eqnarray*}
and $\calI_G=\calK_G\cup \calB_G$. For $\uu\in \ZZ{d}{n}$, we use $\calC_\uu$ to denote
$\{\hspace{0.05cm}\rr\in \ZZ{d}{4n+2},\|\rr-\Psi(\uu)\|\le 2\hspace{0.05cm}\}$.
As the local structure of $\calB_G\cap \calC_\uu$
  depends only on $\mathbb{B}_G(\uu)$, we introduce the following definitions.
\begin{defi}[\mbox{\rm Local Kernel and Boundary}]
For each pair $\pi=(\ss_1,\ss_2)\in \calS^d$ with $d\ge 2$,
let $\calK_{d,\pi}$ and $\calB_{d,\pi}$ be two subsets of $\ZZ{d}{[-2,2]}=\{-2,-1,0,1,2\}^d$,
  such that\vspace{-0.08cm}
\begin{enumerate}
\item if $\ss_1=\ss_2=\text{``no''}$, then $\calK_{d,\pi}=\calB_{d,\pi}=\emptyset$;\vspace{-0.14cm}
\item if $\ss_1=\text{``no''}$ and $\ss_2\not=\text{``no''}$
  \emph{(}$\ss_2=\ee_d$\emph{)},
   then $\calK_{d,\pi}= \{-\ee_d,\00,\ee_d,2\ee_d\}$ and\vspace{-0.05cm}
$$\calB_{d,\pi}=\big\{\hspace{0.08cm}\rr\in \ZZ{d}{[-2,2]}-\calK_{d,\pi}
  \hspace{0.1cm}\big|\hspace{0.1cm}
  \exists\ \rr'\in \{\00,\ee_d,2\ee_d\hspace{0.06cm}\},\|\rr-\rr'\|=1\hspace{0.06cm}\big\};\vspace{-0.2cm}$$
\item if $\ss_1\not=\text{``no''}$ and $\ss_2 =\text{``no''}$
   \emph{(}$\ss_1=\ee_d$\emph{)}, then
  $\calK_{d,\pi}= \{-2\ee_d,-\ee_d,\00,\ee_d\}$ and\vspace{-0.05cm}
$$\calB_{d,\pi}=\big\{\hspace{0.08cm}\rr\in \ZZ{d}{[-2,2]}-\calK_{d,\pi}
  \hspace{0.1cm}\big|\hspace{0.1cm}\exists\
   \rr'\in \{-2\ee_d,-\ee_d,\00\},\|\rr-\rr'\|=1\hspace{0.06cm}\big\};\vspace{-0.2cm}$$
\item otherwise, $\calK_{d,\pi}=\{-2\ss_1,-\ss_1,0,\ss_2,2\ss_2\}$ and\vspace{-0.05cm}
$$\calB_{d,\pi}=\big\{\hspace{0.08cm}\rr\in \ZZ{d}{[-2,2]}-
  \calK_{d,\pi}\hspace{0.1cm}\big|\hspace{0.1cm}\exists\ \rr'\in
  \calK_{d,\pi},\|\rr-\rr'\|=1\hspace{0.06cm}\big\}.\vspace{-0.2cm} $$
\end{enumerate}
\end{defi}
For $\rr\in \mathbb{Z}^d$ and set $S \subseteq \mathbb{Z}^d$,
   let $\rr+S = \{\hspace{0.04cm}\rr+\rr',\rr'\in S\hspace{0.04cm}\}$.
We will use the fact that for all $\uu\in \ZZ{d}{n}$, if $\pi=\BBO_G(\uu)$,
  then $\calK_G\cap \calC_\uu=\Psi(\uu)+\calK_{d,\pi}$ and
  $\calB_G\cap \calC_\uu=\Psi(\uu)+\calB_{d,\pi}$.

\subsubsection*{The Construction of Brouwer Function}

\begin{figure}[!t]
\center
\includegraphics[height=2cm]{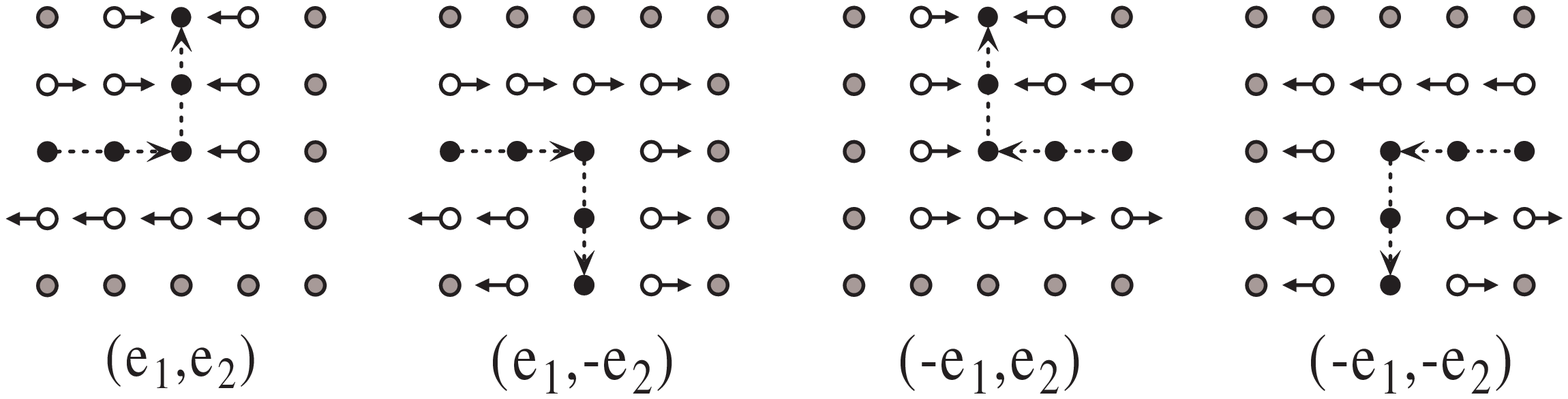}
\includegraphics[height=2cm]{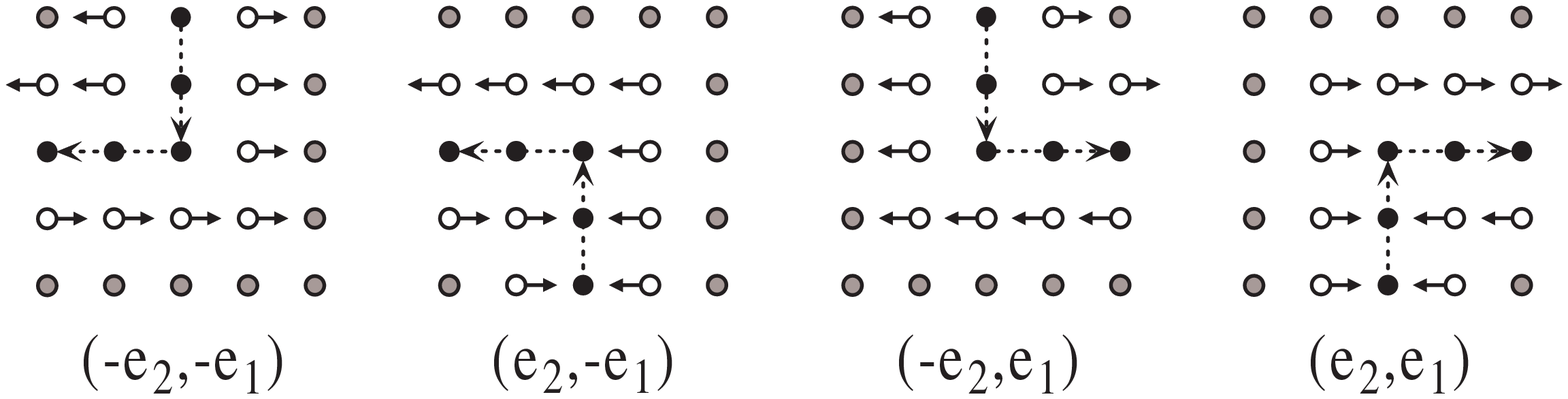}
\includegraphics[height=2cm]{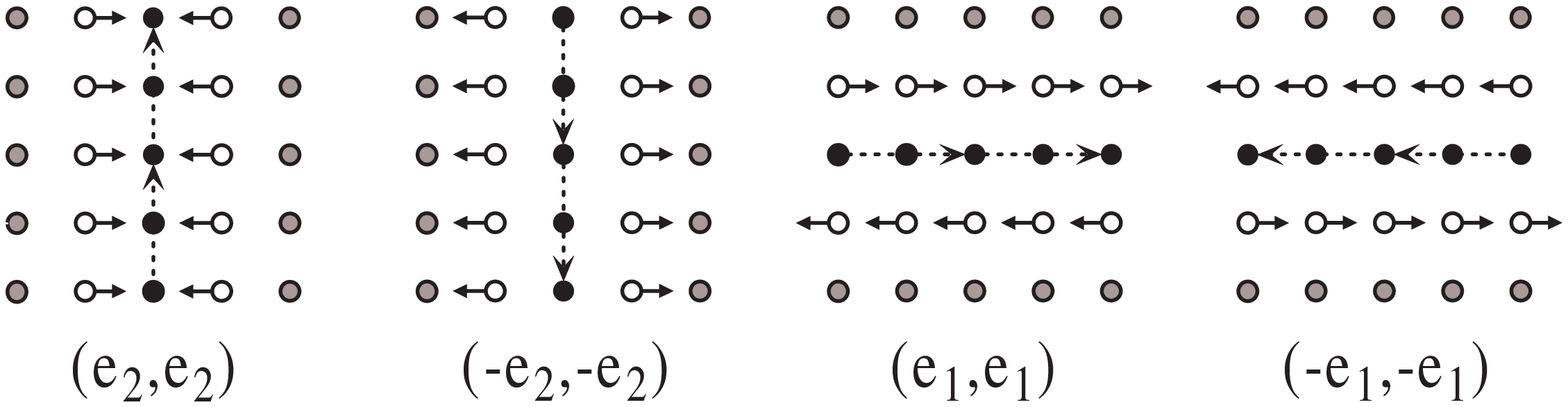}
\includegraphics[height=2cm]{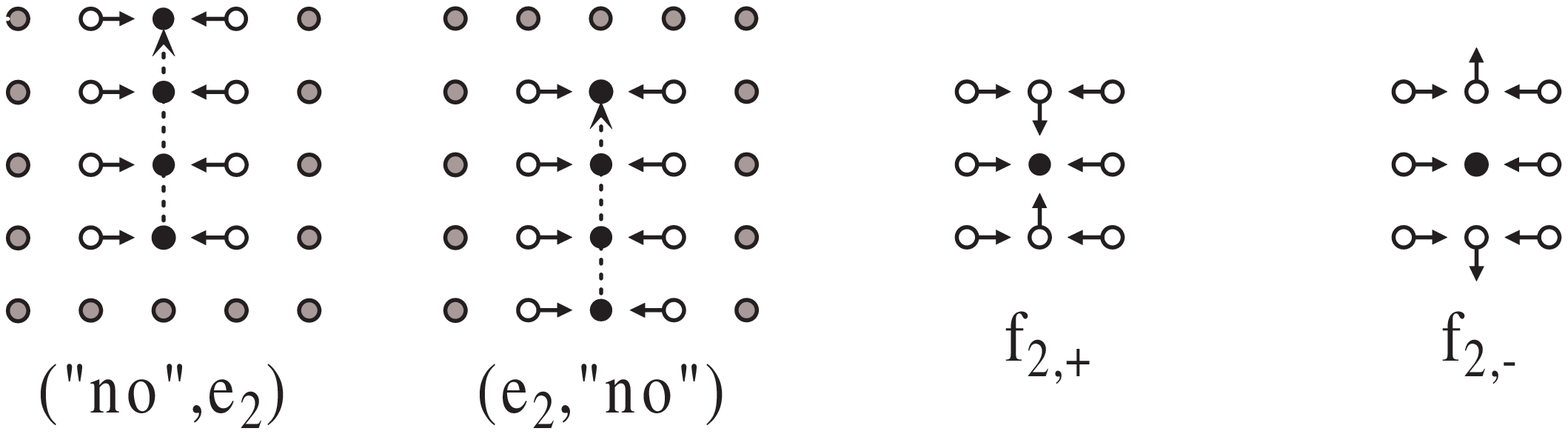}\vspace{-0.1cm}
\caption{$f_{2,\pi}$, $f_{2,+}$ and $f_{2,-}$}\label{fig2doriginal}\vspace{-0.2cm}
\end{figure}

First, we define two direction-preserving functions
  $f_{d,+},f_{d,-}$ from $\calB_d$ to $\bbE^d$ for $d\ge 2$,
  where $\calB_d = \{\hspace{0.04cm}-1,0,1\hspace{0.04cm}\}^d-\00$:
  For every $\rr\in \calB_d$, letting $k$ be the smallest integer such that
  $r_k\not=0$, $f_{d,+}(\rr)=f_{d,-}(\rr)=-r_k\ee_k$
  if $1\le k\le d-1$ and  $f_{d,+}(\rr)=-r_d\ee_d$, $f_{d,-}(\rr)=r_d\ee_d$,
  otherwise.
Using these two functions,
  we inductively build a (\hspace{0.04cm}direction-preserving\hspace{0.04cm})
  function $f_{d,\pi}$ on
  $\calB_{d,\pi}$ for each $\pi=(\ss_1,\ss_2)\in \calS^d$.
See Figure~\ref{fig2doriginal} for the complete construction for $d=2$.
Informally, if $\rr\in \calB_{2,\pi}$ is on the left side
   of the ``local'' path, then $f_{2,\pi}(\rr)=-\ee_1$,
   otherwise it equals $\ee_1$.\vspace{0.025cm}

For $d\ge 3$, the construction is more complex but
  relatively procedural\footnote{Sorry for so many cases.
You will find that they are progressively easier to understand.}.
Below, we use $D$ to denote the map
  $D(\rr)=(r_1,r_2,...,r_{d-1})$ from $\mathbb{Z}^d$ to $\mathbb{Z}^{d-1}$ and
   $U_k$ to denote the map $U_k(\rr)=(r_1,r_2,...,r_{d-1},k)$
   from $\mathbb{Z}^{d-1}$ to $\mathbb{Z}^d$;
we also extends it to sets, that is, $U_k(S) = \{\hspace{0.04cm}U_k(\rr),\rr\in S\hspace{0.04cm}\}$
  for $S\subset \mathbb{Z}^{d-1}$.
Let $S[k] = \{\hspace{0.04cm}\rr\in S,r_d=k\hspace{0.04cm}\}$ for $S\subset \mathbb{Z}^d$.\vspace{0.04cm}
\begin{enumerate}
\item {\bf Moving within (d-1)-dimensional space:}

When $\pi = (\ss_1,\ss_2)$ satisfies
    $\pi'=(D(\ss_1),D(\ss_2))\in \calS^{d-1}$,
    $\calB_{d,\pi}$ can be decomposed into\vspace{-0.05cm}
$$\left(U_{-1}(\calB_{d-1,\pi'})\hspace{0.04cm}\cup
  \hspace{0.04cm}U_{-1}(\calK_{d-1, \pi'})\right)
  \hspace{0.06cm}\cup\hspace{0.06cm} U_0(\calB_{d-1,\pi'})
  \hspace{0.06cm}\cup\hspace{0.06cm} \left(U_{1}(\calB_{d-1,\pi'})
  \hspace{0.04cm}\cup\hspace{0.04cm} U_{1}(\calK_{d-1,\pi'})\right).\vspace{-0.05cm}$$
We set $f_{d,\pi}(\rr)=U_0(f_{d-1,\pi' }(D(\rr)))$
  for $\rr\in U_{-1}(\calB_{d-1,\pi'})\cup U_0(\calB_{d-1,\pi'})
  \cup U_{1}(\calB_{d-1,\pi'})$.\vspace{-0.08cm}

We set $f_{d,\pi}(\rr)=-\ee_{d-1}$ for $\rr\in U_{-1}(\calK_{d-1,\pi'})$
  and $f_{d,\pi}(\rr)=\ee_{d-1}$ for $\rr\in U_1(\calK_{d-1,\pi'})$. \vspace{0.1cm}

\item {\bf Moving along $\pm \ee_d$:} In this case, we will use the fact
 $D(\rr)\in \calB_{d-1}$, for all $\rr\in \calB_{d,\pi}$.

When $\pi=(\ee_d,\ee_d)$, $(\text{``no''},\ee_d)$ or
  $(\ee_d,\text{``no''})$, $f_{d,\pi}(\rr)=U_0(f_{d-1,+}(D(\rr)))$ for all $\rr\in
  \calB_{d,\pi}$.\vspace{-0.08cm}

When $\pi=(-\ee_d,-\ee_d)$, we set $f_{d,\pi}(\rr)=U_0(f_{d-1,-}(D(\rr)))$
  for all $\rr\in \calB_{d,\pi}$.\vspace{-0.05cm}

\item {\bf Moving between $\ee_{d-1}$ and $\pm\ee_d$}:
\begin{enumerate}
\item
When $\pi =(\ss_1,\ss_2)=(\ee_{d-1},\ee_d)$,
   let $\pi'=(D(\ss_1),\text{``no''})\in \calS^{d-1}$. We have
   \begin{eqnarray*}
   &\calB_{d,\pi}[2] =U_2(\calB_{d-1}),\
   \calB_{d,\pi}[1]=U_1(\calB_{d-1,\pi'}\cup \calK_{d-1,\pi'})-\{\ee_d\},
   \calB_{d,\pi}[-2]= \emptyset,&\\
   &\calB_{d,\pi}[0]=U_0(\calB_{d-1,\pi'})\cup\{\ee_{d-1}\},\
   \calB_{d,\pi}[-1]=U_{-1}(\calB_{d-1,\pi'}\cup \calK_{d-1,\pi'}).&
   \end{eqnarray*}
We set $f_{d,\pi}(\rr)=U_0(f_{d-1,+}(D(\rr)))$ for $\rr\in \calB_{d,\pi}[2]$;

$f_{d,\pi}(\rr)=U_0(f_{d-1,\pi'}(D(\rr)))$ for $\rr\in U_{1}(\calB_{d-1,\pi'})
  \cup U_0(\calB_{d-1,\pi'})\cup U_{-1}(\calB_{d-1,\pi'})$;

$f_{d,\pi}(\ee_{d-1}+\ee_d)= f_{d,\pi}(\ee_{d-1})=-\ee_{d-1}$;
$f_{d,\pi}(\rr)=-\ee_{d-1}$ for $\rr\in U_{-1}(\calK_{d-1,\pi'})$;

$f_{d,\pi}(\rr)=\ee_{d-1}$ for
  $\rr\in U_{1}(\calK_{d-1,\pi'})-\{\ee_d,\ee_{d-1}+\ee_d\}$.\vspace{0.08cm}

\item
When $\pi=(\ss_1,\ss_2)=(-\ee_d,\ee_{d-1})$,
   let $\pi'=(\text{``no''},D(\ss_2))\in \calS^{d-1}$. We have
\begin{eqnarray*}
  &\calB_{d,\pi}[-2]=\emptyset,\ \calB_{d,\pi}[2]=U_2(\calB_{d-1}),\
  \calB_{d,\pi}[1]=U_1(\calB_{d-1,\pi'}\cup \calK_{d-1,\pi'})-\{\ee_d\},&\\
  &\calB_{d,\pi}[0]=U_0(\calB_{d-1,\pi'})\cup\{-\ee_{d-1}\},\
  \calB_{d,\pi}[-1]=U_{-1}(\calB_{d-1,\pi'}\cup \calK_{d-1,\pi'}).&
\end{eqnarray*}
We set $f_{d,\pi} (\rr ) =U_0(f_{d-1,-}(D(\rr)))$ for $\rr\in \calB_{d,\pi}[2]$;

$f_{d,\pi}(\rr)=U_0(f_{d-1,\pi'}(D(\rr)))$ for
  $\rr\in U_{1}(\calB_{d-1,\pi'})\cup U_0(\calB_{d-1,\pi'})
   \cup U_{-1}(\calB_{d-1,\pi'})$;

$f_{d,\pi}(\rr)=-\ee_{d-1}$
  for $\rr\in U_{-1}(\calK_{d-1,\pi'})$;
$f_{d,\pi}(-\ee_{d-1}+\ee_d)=f_{d,\pi}(-\ee_{d-1})=-\ee_{d-1}$;

$f_{d,\pi}(\rr)=\ee_{d-1}$ for $\rr\in U_1(\calK_{d-1,\pi'})-\{-\ee_{d-1}+\ee_d,\ee_d\}$.\vspace{0.14cm}

\begin{figure}[!t]
\begin{center}
\includegraphics[width=3.8cm]{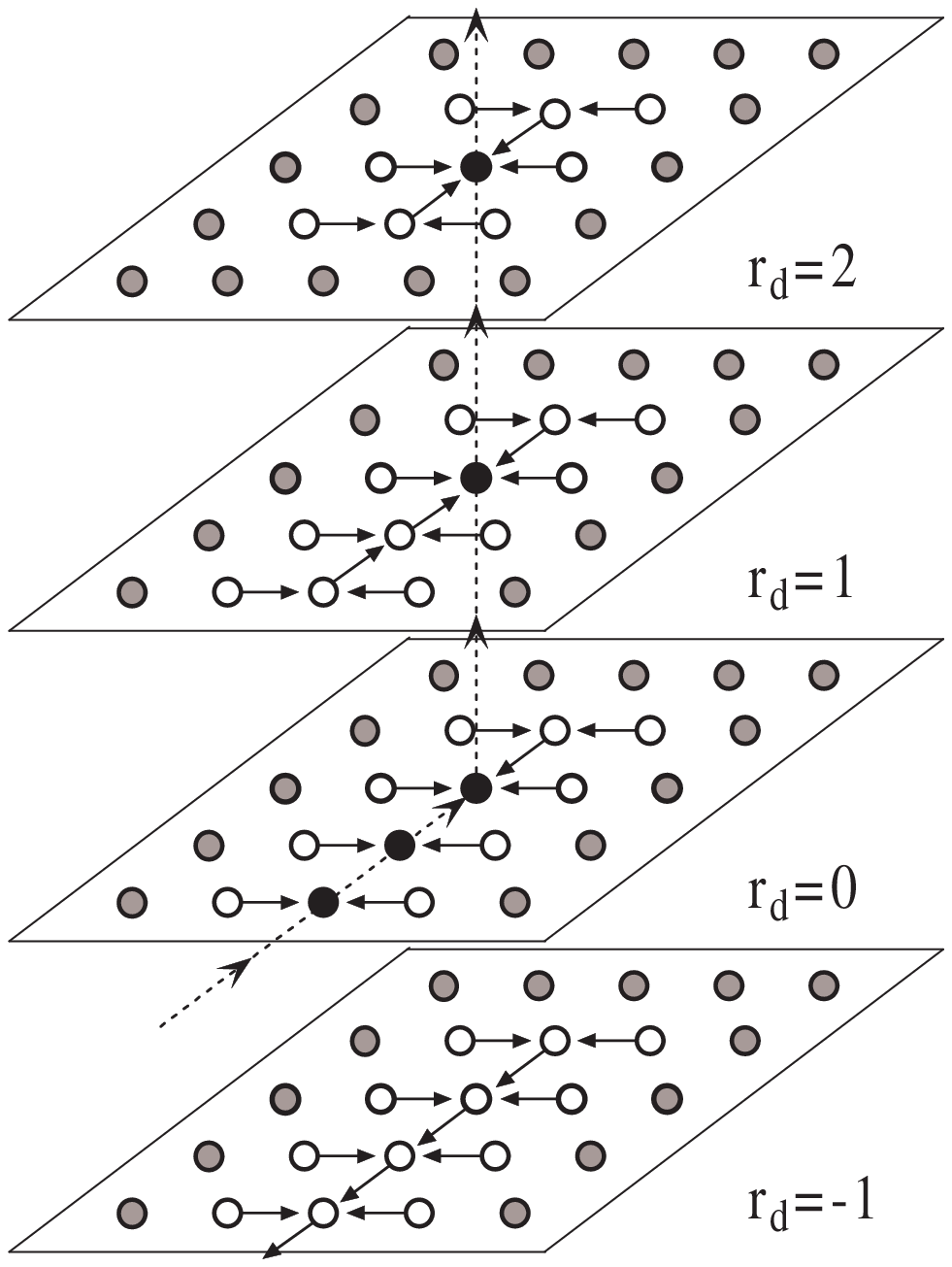}
\includegraphics[width=3.8cm]{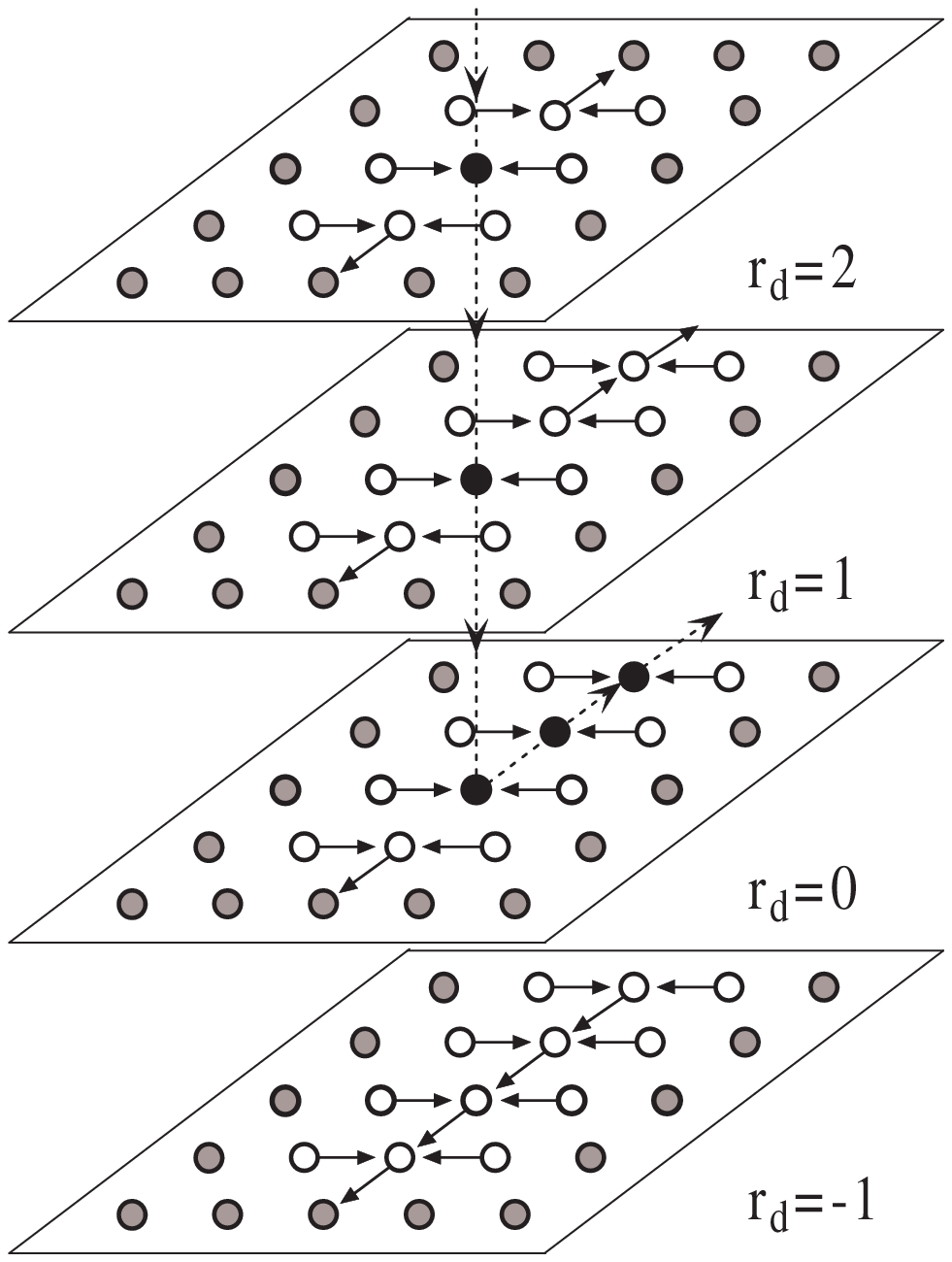}
\includegraphics[width=3.8cm]{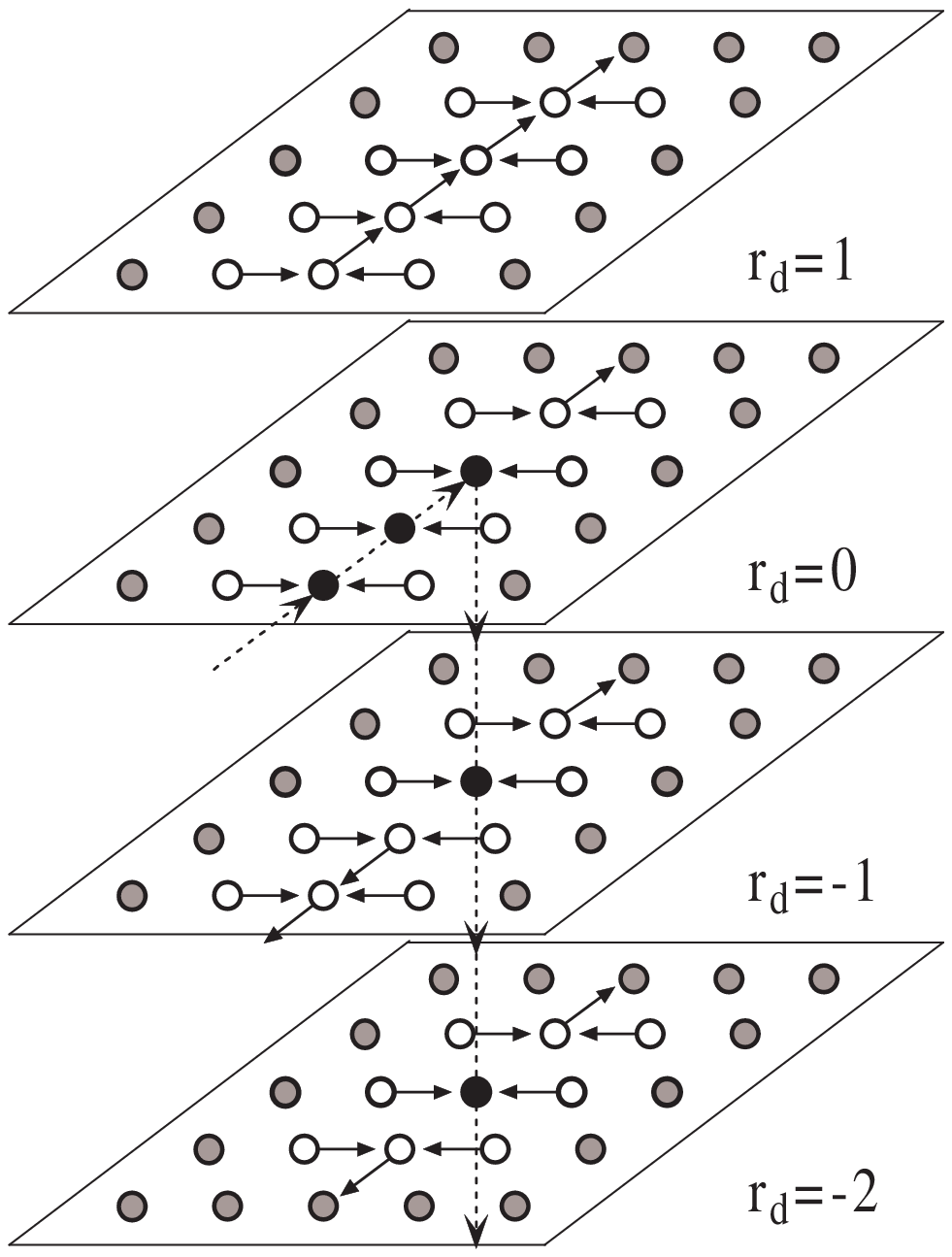}
\includegraphics[width=3.8cm]{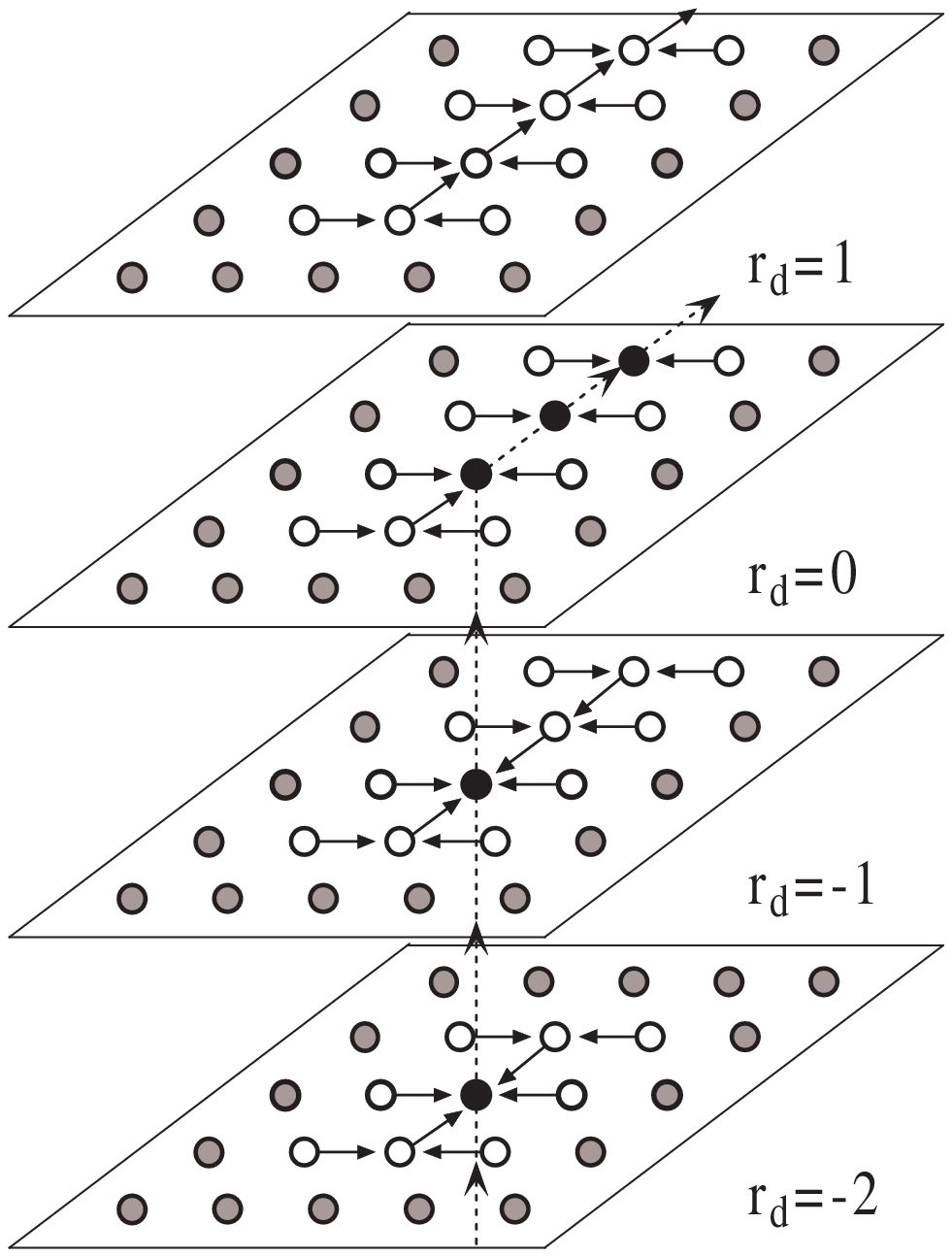}\vspace{-0.25cm}
\end{center}
\caption{$f_{3,\pi}$ where
$\pi=(\ee_{d-1},\ee_d),(-\ee_d,\ee_{d-1}),(\ee_{d-1},-\ee_d)$ and $(\ee_d,\ee_{d-1})$}\label{fig3d1}\vspace{-0.1cm}
\end{figure}


\item
When $\pi=(\ss_1,\ss_2)=(\ee_{d-1},-\ee_d)$, let
   $\pi'=(D(\ss_1),\text{``no''})\in \calS^{d-1}$.
We have
\begin{eqnarray*}
   &\calB_{d,\pi}[2] =\emptyset,\
   \calB_{d,\pi}[1]=U_1(\calB_{d-1,\pi'}\cup \calK_{d-1,\pi'}),\
   \calB_{d,\pi}[0]=U_0(\calB_{d-1,\pi'})\cup\{\ee_{d-1}\},&\\
   &\calB_{d,\pi}[-1]=U_{-1}(\calB_{d-1,\pi'}\cup \calK_{d-1,\pi'})-\{-\ee_d\},\
   \calB_{d,\pi}[-2]=U_{-2}(\calB_{d-1}).&
\end{eqnarray*}
We set $f_{d,\pi} (\rr ) =U_0(f_{d-1,-}(D(\rr)))$ for $\rr\in \calB_{d,\pi}[-2]$;

$f_{d,\pi}(\rr)=U_0(f_{d-1,\pi'}(D(\rr)))$
  for $\rr\in U_{1}(\calB_{d-1,\pi'})\cup U_0(\calB_{d-1,\pi'})\cup U_{-1}(\calB_{d-1,\pi'})$;

$f_{d,\pi}(\rr)=\ee_{d-1}$ for $\rr\in U_{1}(\calK_{d-1,\pi'})$;
$f_{d,\pi}(\ee_{d-1})=f_{d,\pi}(\ee_{d-1}-\ee_d)=\ee_{d-1}$;

$f_{d,\pi}(\rr)=-\ee_{d-1}$ for
  $\rr\in U_{-1}(\calK_{d-1,\pi'})-\{-\ee_d,\ee_{d-1}-\ee_d\}$.\vspace{0.14cm}

\item
When $\pi=(\ss_1,\ss_2)=(\ee_d,\ee_{d-1})$,
   let $\pi'=(\text{``no''},D(\ss_2))\in \calS^{d-1}$.
We have
\begin{eqnarray*}
  &\calB_{d,\pi}[2]=\emptyset,\
  \calB_{d,\pi}[1]=U_1(\calB_{d-1,\pi'}\cup \calK_{d-1,\pi'}),\
  \calB_{d,\pi}[0]=U_0(\calB_{d-1,\pi'})\cup\{-\ee_{d-1}\},&\\
  &\calB_{d,\pi}[-1]=U_{-1}(\calB_{d-1,\pi'}\cup \calK_{d-1,\pi'})-\{-\ee_d\},\
  \calB_{d,\pi}[-2]=U_{-2}(\calB_{d-1}).&
\end{eqnarray*}
We set $f_{d,\pi} (\rr ) =U_0(f_{d-1,+}(D(\rr)))$ for $\rr\in \calB_{d,\pi}[-2]$;

$f_{d,\pi}(\rr)=U_0(f_{d-1,\pi'}(D(\rr)))$ for
  $\rr\in U_{1}(\calB_{d-1,\pi'})\cup U_0(\calB_{d-1,\pi'})\cup U_{-1}(\calB_{d-1,\pi'})$;

$f_{d,\pi}(\rr)=\ee_{d-1}$ for $\rr\in U_{1}(\calK_{d-1,\pi'})$;
$f_{d,\pi}(-\ee_{d-1})=f_{d,\pi}(-\ee_{d-1}-\ee_d)=\ee_{d-1}$;

$f_{d,\pi}(\rr)=-\ee_{d-1}$ for
$\rr\in U_{-1}(\calK_{d-1,\pi'}))-\{-\ee_d,-\ee_{d-1}-\ee_d\}$.\vspace{0.1cm}
\end{enumerate}
\end{enumerate}

\begin{lemm}[\mbox{\rm Locally Directional Preserving}]\label{coro1}
For every $\pi\in \calS^d$, $f_{d,\pi}$ is
  direction-preserving on $\calB_{d,\pi}$.
\end{lemm}
\begin{proof}
We prove the lemma by induction on $d$. The base case when $d=2$
  is trivial. We now consider the case when $d>2$ and assume inductively
  that the statement is true for $d-1$.

First, $\pi=(\ee_d,\ee_d)$, $(\text{``no''},\ee_d)$, $(\ee_d,\text{``no''})$
  or $(-\ee_d,-\ee_d)$. The statement follows
  from the fact that $f_{d-1,+}$ and $f_{d-1,-}$ are direction-preserving
  on $\calB_{d-1}$.
Second, $\pi=(\ss_1,\ss_2)$ satisfies $\pi'=(D(\ss_1),D(\ss_2))\in \calS^{d-1}$.
  By the inductive hypothesis, $f_{d-1,\pi'}$ is direction-preserving, from
  which the statement follows.
Third, $\pi=(\ee_{d-1},\ee_d)$, $(\ee_{d-1},-\ee_d)$, $(-\ee_d,\ee_{d-1})$ or
  $(\ee_d,\ee_{d-1})$. One can prove the following statement by induction on $d$.\vspace{-0.05cm}
\begin{quote}
For $\pi_1=(\text{``no''},\ee_d)$ and $\pi_2=(\ee_d,\text{``no''})$,
$\calB_{d,\pi_1}\cap \calB_d=\calB_{d,\pi_2}\cap
\calB_d=\calB_d-\{-\ee_d,\ee_d\}$.
Moreover for each $\rr\in \calB_d-\{-\ee_d,\ee_d\}$, $f_{d,\pi_1}(\rr)
=f_{d,\pi_2}(\rr)=f_{d,+}(\rr)=f_{d,-}(\rr)$.\vspace{-0.05cm}
\end{quote}
\indent To show $f_{d,\pi}$ is direction-preserving on $\calB_{d,\pi}$, it suffices to check
   $\|\rr_1-\rr_2\|>1$, for all pairs $\rr_1,\rr_2\in \calB_{d,\pi}$
   such that$f_{d,\pi}(\rr_1)=\ee_{d-1}$
   and $f_{d,\pi}(\rr_2)=-\ee_{d-1}$.
\end{proof}

With these local functions $f_{d,\pi}$,
  we can  build a global function $f^G$ from $\calB_G$
  to $\{\pm \ee_1,...,$ $\pm \ee_{d-1}\}$ as following:
for every $\rr\in \calB_G$, we set
  $f^{G}(\rr)=f_{d,\pi}(\rr-\Psi(\uu))$,
  where $\uu$ is the   lexicographically smallest vertex in $\ZZ{d}{n}$
  such that $\rr\in \calC_\uu$ and $\pi=\mathbb{B}_G(\uu)$,

\begin{figure}[tb]

\rule{\textwidth}{1pt}\vspace{0.06cm}

\textbf{\ \ $f_{G}(\rr)$, where $\rr\in \ZZ{d}{4n+2}$}

\vspace{-0.16cm}\rule{\textwidth}{1pt}\vspace{0.2cm}

\begin{tabular}{@{\hspace{0.1cm}}r@{\hspace{0.2cm}}p{\textwidth}}
\ \ 1\hspace{0.05cm}: & let $\uu^*$ and $\ww^*$ be the starting and
   ending vertices of $G$, $\pp^*=\Psi(\uu^*)$ and $\qq^*=\Psi(\ww^*)$\\[0.5ex]

\ \ 2\hspace{0.05cm}: &
  \textbf{if} $r_d=1$ or $2$, and $D(\rr)=D(\pp^*)$ \textbf{then} $f_{G}(\rr)=\ee_d$ \\[0.5ex]

\ \ 3\hspace{0.05cm}: & \textbf{else if} $r_d=1$ or $2$, and $\|D(\rr)-D(\pp^*)\|=1$ \textbf{then} \\[0.5ex]

\ \ 4\hspace{0.05cm}: & \ \ \ \ let $k$ denote the smallest integer such that
$r_k\not=r^*_k$, $f_G(\rr)=(p^*_k-r_k)\ee_k$ \\[0.5ex]

\ \ 5\hspace{0.05cm}: & \textbf{else if} $\rr=\qq^*$ \textbf{then} $f_G(\rr)=0$ \\[0.5ex]

\ \ 6\hspace{0.05cm}: & \textbf{else if} $\rr=\qq^*+\ee_d$ \textbf{then} $f_G(\rr)=-\ee_d$ \\[0.5ex]

\ \ 7\hspace{0.05cm}: & \textbf{else if} $\rr\in \calK_G$ \textbf{then} $f_G(\rr)=\ee_d$ \\[0.5ex]

\ \ 8\hspace{0.05cm}: & \textbf{else if} $\rr\in \calB_G$ \textbf{then} $f_G(\rr)=f^G(\rr)$ \\[0.5ex]

\ \ 9\hspace{0.05cm}: & \textbf{else if} $r_d=1$ (and $\|D(\rr)-D(\pp^*)\|\ge 2$) \textbf{then} \\[0.5ex]

\ \ 10\hspace{0.05cm}: & \ \ \ \ let $k$ denote the smallest integer such that
$r_k\not=r^*_k$, $f_G(\rr)=\text{sign}(p^*_k-r_k)\ee_k$ \\[0.5ex]

\ \ 11\hspace{0.05cm}: & \textbf{else} $f_G(\rr)=-\ee_d$\\[0.6ex]
\end{tabular}

\vspace{0.097cm}\rule{\textwidth}{1pt} \caption{Construction of Function $f_{G}$ from $f^G$}\label{ext}
\end{figure}

\begin{lemm}
For every canonical grid PPAD graph $G$ over $\Zdn$,
  $f^G$ is direction-preserving on set $\calB_G$.
\end{lemm}
\begin{proof}
By Lemma~\ref{coro1}, it suffices to prove the following: For
  $\rr\in \calB_G$, if $\rr\in \calC_\uu\cap \calC_\vv$ where
  $\uu,$ $\vv\in \Zdn$, then $f_{d,\pi_1}(\rr-\Psi(\uu))=
  f_{d,\pi_2}(\rr-\Psi(\vv))$,
  where $\pi_1=\mathbb{B}_G(\uu)$ and $\pi_2=\mathbb{B}_G(\vv)$.
We will use the fact that  $\ss=\uu-\vv\in   \mathbb{E}^d$
  and either $(\uu,\vv)\in G$ or $(\vv,\uu)\in G$.

For $S\subset \mathbb{Z}^d$ and $\pp\in\mathbb{Z}^d$,
  we use $S+\pp$ to denote $\{\hspace{0.04cm}\rr\in \mathbb{Z}^d
  \hspace{0.1cm}|\hspace{0.1cm}\rr=\rr'+\pp,\rr'\in S\hspace{0.04cm}\}$.
The lemma is a direct consequence of the following statement
  which can be proved by induction on $d$.\vspace{-0.05cm}
\begin{quote}
For all  $\ss=b\hspace{0.04cm}\ee_k\in \bbE^d$ with
  $b\in \{\pm 1\}$ and  $k \in [1: d]$,
  if $\pi_1 =$ $(\ss_1,\ss)\in \calS^d$
  and $\pi_2=(\ss,\ss_2)\in \calS^d$
  then $\{\hspace{0.06cm}\rr\in \calB_{d,\pi_1},r_k=2b\hspace{0.06cm}\}
  =\{\hspace{0.06cm}\rr\in \calB_{d,\pi_2},r_k=-2b
  \hspace{0.06cm}\}+4\hspace{0.04cm}\ss,$
  and for every $\rr$ in the former set,
  $f_{d,\pi_1}(\rr)=f_{d,\pi_2}(\rr-4\ss)$.\vspace{-0.05cm}
\end{quote}
\vspace{-0.27in}
\end{proof}

Finally, to extend $f^{G}$ onto $\ZZ{d}{4n+2}$
  to define our function $f_G$, we apply the procedure
  given in Fig.~\ref{ext}.
It is somewhat tedious but procedural
  to check that $f_G$ satisfies
  both \textbf{Property D.2} and \textbf{D.1} stated at the beginning
  of this subsection.

\section{Randomized Lower Bound for $\text{\sf ES}^d$}\label{sec:ESLOWER}

The technical objective of this section is to construct a distribution
$\calS $ of
  $d$-non-repeating strings and show that,
  for a random string $S$ drawn according to $\calS $,
  every deterministic algorithm for $\text{\sf ES}^d$ needs expected
  $(\Omega(n))^{d}$ queries to $\BBO_{S}$.
Thus, by Yao's Minimax Principle \cite{Yao77},
   we have $\text{\sf RQ}_{\text{\sf ES}}^{d}(n) = (\Omega (n))^{d}$.
Our main Theorem \ref{thm:FixedPoint} then follows
  from Theorems \ref{thm:EoS2GPPAD} and \ref{thm:GP2ZP}.\vspace{0.005cm}

We apply random permutations
  hierarchically to define distribution $\calS $
  to ensure that
  a random string from $\calS $ has sufficient
  entropy that its search problem
  is expected to be difficult.
The use the hierarchical structure
  guarantees that each string in $\calS $ is
  $d$-non-repeating.

\subsection{Hierarchical Construction of Random $d$-Non-Repeating Strings}

We first define our hierarchical framework.
Let $\mathbb{J}_{n}=[2:2n+2]$, $\mathbb{O}_{n}=\{3,5,...,2n+1\}$ and
  $\mathbb{F}_{n}=\{4,6,...,2n+2\}$.
Let $S_0 = 2$, $S_1=3\circ 4$, ..., $S_n=(2n+1)\circ (2n+2)$.
Each permutation $\pi$ from $[1:n]$ to  $[1:n]$
  defines a string $C = S_{0}\circ S_{\pi (1)}\circ \dotsb \circ S_{\pi
 (n)}$ which we refer to as a {\em connector} over $\mbbJ_{n}$.\vspace{0.015cm}

Let $r[C] = 2\pi (n)+2$, the last symbol of $C$.
We use $\phi_{C } (2)$ to denote the right neighbor of $2$.
Each $s \in \mbbJ_{n} - \{2,r[C_{\pi } ]\}$
   has two neighbors in $C$.
The left neighbor of an even $s$ is $s-1$,
  we  use  $\phi_{C} (s)$ to denote its right neighbor;
the right neighbor of an odd $s$ is
  $s+1$, and we use $\phi_{C } (s)$ to denote its left neighbor.
Clearly, if $\phi_{C } (s) =t$ then  $\phi_{C } (t) =s$.\vspace{0.015cm}

Our hierarchical framework is built on
  $T_{n,d}$, the rooted complete-$(2n+1)$-nary tree of height $d$.
In $T_{n,d}$, each internal node $u$ is connected to its $(2n+1)$ children
  by edges with distinct labels from  $\mathbb{J}_{n}$;
  if $u$ is connected to $v$ by an edge
  labeled with $j$, then we call $v$ the {\em  $j^{{th}}$-successor}
  of $u$.
Each node $v$ of $T_{n,d}$ has a {\em natural name}, $\name{v}$,
  the concatenation of labels along the path from the root of $T_{n,d}$ to $v$.
Let $\height{v}$ and $\level{v}$ denote the height and level of node
  $v$ in the tree.
For example, the height of the root is $d$ and the level of the
  root is $0$.\vspace{0.04cm}

\begin{defi}[\mbox{\rm Tree-of-Connectors}]
An $(n,d)$-{\rm ToC} $\calT $ is a tree $T_{n,d}$ in which each internal node
  $v$ is associated with a connector $C_{v}$ over $\mbbJ_{n}$.
The $r[C_{v}]^{th}$-successor is referred to as the {\rm last successor}
  of $v$.
The {\rm tail} of $v$, $\tail{v}$, is the leaf reachable from $v$
  by last-successor relations.
The tail of a leaf is itself.
The {\rm tail} of $\calT $, $\tail{\calT }$, is the tail of its root.
The {\em head} of a leaf $u$, $\head{u}$,
   is the ancestor of $u$ with the largest height such that $u$ is
   its tail.\vspace{0.04cm}
\end{defi}

\begin{defi}[\mbox{\rm Valid ToC}]
An $(n,d)$-{\rm ToC} $\calT $ is {\rm valid}
   if for all internal $v$ and for each pair of $s,t\in \mbbJ_{n}$ with
  $\phi_{C_{v}} (s) = t $,
   $\name{u_{s}}$ and $\name{u_{t}}$ share a common suffix of length
  $\height{v}-1$, where
   $u_{s}$ and $u_{t}$, respectively, are  the
   tails of the $s^{th}$-successor and $t^{th}$-successor of $v$.
\end{defi}

\begin{figure}[!h]
\begin{center}\vspace{-0.05cm}
\includegraphics[width=14cm]{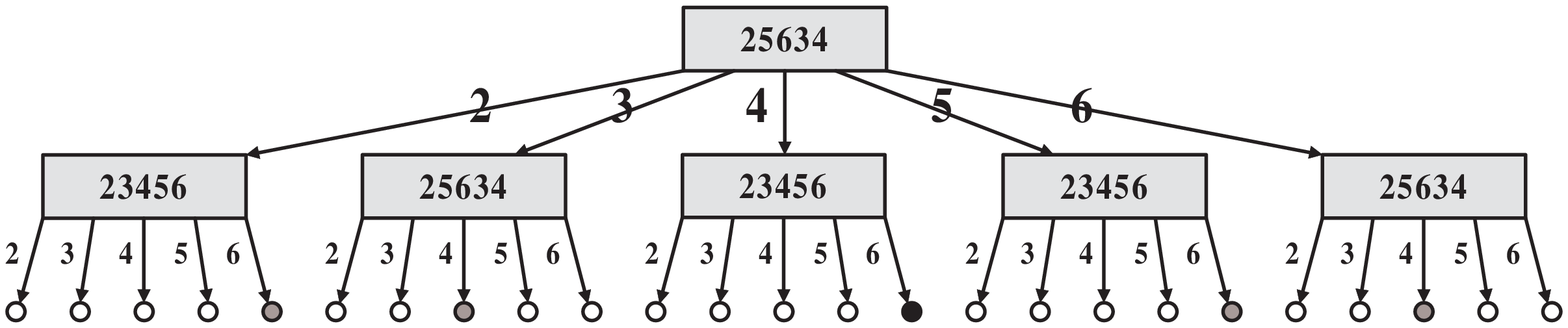}\vspace{-0.35cm}
\end{center}
\caption{A valid $(2,2)$-ToC $\calT$} \label{fig2db}
\end{figure}
\vspace{-0.05in}

\begin{defi}[\mbox{$\BBO_{\calT }$} for  accessing $\calT $]
Suppose $\calT$ is a valid $(n,d)$-ToC.
The input to $\mathbb{B}_\calT$ is a point $\qq$ from $(\mbbJ_{n})^{d}$
  (defining the name of a leaf $u$ in $\calT $).
Let $h = \height{\head{u}}$.
If $u$ is the tail of $\calT $, i.e., $h = d$,
   then $\BBO_{\calT } = \calT $.
Otherwise, let $v_{1} = \head{u}$ and let $v$ be the parent of
   $v_{1}$.
Note that $v_{1}$ is the $q_{d-h}^{th}$-successor of $v$.
Let $\calT_{1}$ be the tree rooted at $v_{1}$.
As $u \neq \tail{v}$, $\phi_{C_{v}}
  (q_{d-h})$ is defined and
let $\calT_{2}$ be the subtree rooted
 the $\phi_{C_{v}} (q_{d-h})^{th}$-successor of $v$.
Then, $\BBO_{\calT } (\qq) = (h,\phi_{C_{v}}
  (q_{d-h}), \calT_{1}, \calT_{2})$.\vspace{0.04cm}
\end{defi}

We now define our final search problem
 {\sf   Name-the-Tail},
   on a valid $(n,d)$-ToC.
The search problem {\sf NT}$^d$ is:
  Given a valid $(n,d)$-ToC $\calT^{*} $ accessible by
  $\BBO_{\calT^{*} }$, find the name of its tail.
We will prove Theorem \ref{thm:NT} in  Section \ref{sub:proof}.
Below, we prove Theorem \ref{thm:NT2ES}
  to reduce $\text{\sf NT}^d$~to~$\text{\sf ES}^d$.

\begin{theo}[Complexity of \mbox{\rm $\text{\sf NT}^d$}]\label{thm:NT}
For all sufficiently large $n$,
$$\text{\sf RQ}_{\text{\sf NT}}^d(n)\hspace{0.06cm}\ge \hspace{0.06cm}
\frac{1}{2} \left(\frac{n}{2\cdot 24^{d}} \right)^{d}.$$
\end{theo}

\begin{theo}[From $\text{\sf NT}^d$ to $\text{\sf ES}^d$]\label{thm:NT2ES}
For all $d\geq 1$,
$\text{\sf RQ}_{\text{\sf NT}}^{d}(n)\le
   \text{\sf RQ}_{\text{\sf ES}}^d(4n+4)$.
\end{theo}

\begin{proof}
We need to build
  a $d$-non-repeating string from a valid $(n,d)$-{\rm
  ToC} $\calT $.
In fact, we will construct two strings $S[\calT]$ and
  $Q[\calT]$ over $\mathbb{Z}_{4n+4}$,
  each has length $\Theta (n^{d})\cdot d$.
 $S[\calT]$ starts with $\ss_d$ and ends with $\calF(\name{\tail{\calT }})\in \mathbb{Z}^d$
  while $Q[\calT]$ starts with $\calF(\name{\tail{\calT }})$ and ends with
  $\ss_d$, where for $\pp \in \mathbb{Z}^{d}$
   $\calF(\pp)=(2p_1,...,2p_{d-1},2p_d-1)$
  and  $\ss_d\in \mathbb{Z}^d$ defined to be
   $s_1=1$, and $\ss_d=(2,...,2,1)$ for $d>1$.\vspace{0.02cm}

For any two strings $S_1=a_1...a_{k}$ and $S_2=b_1...b_t$, let
 $S_1\circ S_2 =a_1...a_kb_1...b_t$.
For $d\geq 1$, if $a_{k-d+i}=b_{i}$ for all $1\le i\le d$,
  then let $S_1\circ_d S_2 =a_1...a_kb_{d+1}...b_t$.
Given a string $S$ over $\mathbb{Z}$ of length $k\cdot d$,
   we write $S$ as $\uu_1...\uu_k$ with $\uu_i\in \mathbb{Z}^d$.
Let $\inser{d}{S}{t} =\uu_1\circ t\circ\uu_2\circ t...\uu_{k-1}\circ t\circ \uu_k$.\vspace{0.015cm}

We use the following recursive procedure. Let $r$ be the root of $\calT $.
Assume $C_{r} = a_1...a_{2n+1}$.
When $d=1$, we set $S[\calT]=1b_1b_2... b_{2n+1}$ and
  $Q[\calT]=b_{2n+1} ...b_2b_11$,
  where $b_i=2a_i-1$. When $d\ge 2$,
\vspace{-0.03cm}
\begin{enumerate}
\item  let $\calT_i$ be the subtree of $\calT$ rooted at the
  $a_i^{th}$-successor of $r$
  and let $\pp_i\in (\mathbb{F}_n)^{d-1}$ be the
  name of the tail of $\calT_i$ given by $\calT_{i}$ (not  by $\calT
$).\vspace{-0.08cm}

\item  for every odd $i \in [1:2n+1]$, set $S'_i =
\inser{d-1}{S[\calT_{i}]}{2a_i}$
  (which starts with $\ss_{d-1}$ and ends with $\calF(\pp_i)$)
  and for every even $i \in[1: 2n+1]$  set $S'_i = \inser{d-1}{Q[\calT_{i}]}{2a_i}$
  (which starts with   $\calF(\pp_i)$ and ends with $\ss_{d-1}$);
$S[\calT]
=\ss_d \circ_{d-1} S_1'\circ_{d-1} S_2' \circ_{d-1} S_3'\circ_{d-1}...
  \circ_{d-1} S_{2n}'\circ_{d-1} S_{2n+1}' $;\vspace{-0.07cm}

\item  for every odd $i\in[1: 2n+1]$, set $Q'_i = \inser{d-1}{Q[\calT_{i}]}{2a_i}$
  (which starts with $\calF(\pp_i)$ and ends with $\ss_{d-1}$) and
  for every even $i\in[1: 2n+1]$, set $Q'_i = \inser{d-1}{S[\calT_{i}]}{2a_i}$ (which
starts with $\ss_{d-1}$ and ends with $\calF(\pp_i)$);
$ Q[\calT]=(2a_{2n+1})\circ Q'_{2n+1}\circ_{d-1}
  Q'_{2n}\circ_{d-1}...\circ_{d-1} Q'_2\circ_{d-1} Q'_1\circ \ss_d.$\vspace{0.05cm}
\end{enumerate}

The two strings for the example in Figure \ref{fig2db} above are:\vspace{0.18cm}
{\footnotesize \begin{eqnarray*}
&\hspace{-0.2cm}S[\calT]=\raisebox{0.8ex}{2}\hspace{0.02cm}1
\hspace{0.02cm}\raisebox{0.8ex}{4}\hspace{0.02cm}3\hspace{0.02cm}
\raisebox{0.8ex}{4}\hspace{0.02cm}5\hspace{0.02cm}\raisebox{0.8ex }{4}\hspace{0.02cm}
7\hspace{0.02cm}\raisebox{0.8ex}{4}\hspace{0.02cm}9\hspace{0.02cm}\raisebox{0.8ex}{4}\hspace{0.02cm}11\hspace{0.02cm}
\raisebox{0.8ex}{10}\hspace{0.02cm}9\hspace{0.02cm}\raisebox{0.8ex}{10}\hspace{ 0.02cm}
7\hspace{0.02cm}\raisebox{0.8ex}{10}\hspace{0.02cm}5\hspace{0.02cm}\raisebox{0.8ex}{10}\hspace{0.02cm}3\hspace{0.02cm}
\raisebox{0.8ex}{10}\hspace{0.02cm}1\hspace{0.02cm}\raisebox{0.8ex}{12}\hspace{0.02cm}
3\hspace{0.02cm}\raisebox{0.8ex}{12}\hspace{0.02cm}9\hspace{0.02cm}\raisebox{0.8ex}{12}\hspace{0.02cm}11\hspace{0.02cm}
\raisebox{0.8ex}{12}\hspace{0.02cm}5\hspace{0.02cm}\raisebox{0.8ex}{12}\hspace{0.02cm} 7\hspace{
0.02cm}\raisebox{0.8ex}{6}\hspace{0.02cm}5\hspace{0.02cm}\raisebox{0.8ex}{6}\hspace{0.02cm}11\hspace{0.02cm}
\raisebox{0.8ex}{6}\hspace{0.02cm}9\hspace{0.02cm}\raisebox{0.8ex}{6}\hspace{0.02cm} 3\hspace{0.02cm}\raisebox{
0.8ex}{6}\hspace{0.02cm}1\hspace{0.02cm}\raisebox{0.8ex}{8}\hspace{0.02cm}3\hspace{0.02cm}
\raisebox{0.8ex}{8}\hspace{0.02cm}5\hspace{0.02cm}\raisebox{0.8ex}{8}\hspace{0.02cm}
7\hspace{0.02cm}\raisebox{0.8ex}{8}\hspace{ 0.02cm}9\hspace{0.02cm}\raisebox{0.8ex}{8}\hspace{0.03cm}11\\[0.9ex]
&\hspace{-0.2cm}Q[\calT]=\raisebox{0.8ex}{8}\hspace{0.02cm}11\hspace{0.02cm}\raisebox{0.8ex}{8}\hspace{0.02cm}9\hspace{0.02cm}
\raisebox{0.8ex }{8}\hspace{ 0.02cm}7\hspace{0.02cm}\raisebox{0.8ex}{8}\hspace{0.02cm}
5\hspace{0.02cm}\raisebox{0.8ex}{8}\hspace{0.02cm}3\hspace{0.02cm}\raisebox{0.8ex}{8}\hspace{0.02cm}1\hspace{0.02cm}
\raisebox{0.8ex}{6}\hspace{ 0.02cm}3\hspace{ 0.02cm}\raisebox{0.6ex}{6}\hspace{0.02cm}
9\hspace{0.02cm}\raisebox{0.8ex}{6}\hspace{0.02cm}11\hspace{0.02cm}\raisebox{0.8ex}{6}\hspace{0.02cm}5\hspace{0.02cm}
\raisebox{0.8ex}{6}\hspace{0.02cm}7\hspace{ 0.02cm}\raisebox{ 0.8ex}{12}\hspace{0.02cm}
5\hspace{0.02cm}\raisebox{0.8ex}{12}\hspace{0.02cm}11\hspace{0.02cm}\raisebox{0.8ex}{12}\hspace{0.02cm}9\hspace{0.02cm}
\raisebox{0.8ex}{12}\hspace{0.02cm}3\hspace{0.02cm}\raisebox{ 0.8ex}{12}\hspace{ 0.02cm}
1\hspace{0.02cm}\raisebox{0.8ex}{10}\hspace{0.02cm}3\hspace{0.02cm}\raisebox{0.8ex}{10}\hspace{0.02cm}5\hspace{0.02cm}
\raisebox{0.8ex}{10}\hspace{0.02cm}7\hspace{0.02cm}\raisebox{0.8ex}{10}\hspace{ 0.02cm}
9\hspace{0.02cm}\raisebox{0.8ex}{10}\hspace{0.02cm}11\hspace{0.02cm}\raisebox{0.8ex}{4}\hspace{0.02cm}9\hspace{0.02cm}
\raisebox{0.8ex}{4}\hspace{0.02cm}7\hspace{0.02cm}\raisebox{0.8ex}{4}\hspace{0.02cm} 5\hspace{ 0.02cm
}\raisebox{0.8ex}{4}\hspace{0.02cm}3\hspace{0.02cm}\raisebox{0.8ex}{4}\hspace{0.03cm}1\hspace{0.03cm}
\raisebox{0.8ex}{2}\hspace{0.03cm}1
\end{eqnarray*}
}

\vspace{-0.18in}
The correctness of our construction can be established using the next
two lemmas.
\end{proof}

\begin{lemm}[\mbox{\rm Non-Repeating}]
If $\calT $ is a valid $(n,d)$-{\rm ToC}, then
  both $S[\calT]$ and $Q[\calT]$ are $d$-non-repeating.
\end{lemm}

\begin{proof}
We prove the lemma by induction on $d$. The base case when
  $d=1$ is trivial.

Assume $d>1$ and also inductively that the statement is true for $d-1$.
Suppose for the sake of contradiction that $S'=a_1a_2...a_d\in
  \mathbb{Z}_{4n+4}^d$ appears in
  $S[\calT]$ more than once.
Note that exactly one symbol in $S'$, say $a_t$, is odd.
Let $k\in [1:d]$ be the following integer: if $t=d$, then
  $k=1$; otherwise, $k=t+1$.

First, if $a_k=2$, then $S'$ appears in $S[\calT]$ implies
  $(a_1,a_2,...,a_d)=\ss_d$. But such $S'$ only appears in
  $S[\calT]$ once, which contradicts with the assumption.

Otherwise, let $S''\in \mathbb{Z}_{4n+4}^{d-1}$ be the string obtained
  by removing $a_k$ from $S'$, then: if $a_k/2$ is odd, then $S''$
  appears in $Q[\calT_{a_k/2}]$ more than once; otherwise, $S''$
  appears in $S[\calT_{a_k/2}]$ more than once,
  which contradicts with the inductive hypothesis.

The proof for string $Q[\calT]$ is similar.
\end{proof}

\begin{lemm}[\mbox{\rm Asking $\BBO_{\calT }$}]\label{lem:TocString}
Suppose $\calT$ is a valid $(n,d)$-ToC and $S = S[\calT ]$ and $Q =
  Q[\calT]$.
For any $\uu \in \ZZ{d}{4n+4}$, we can compute $\BBO_{S} (\uu )$ and
  $\BBO_{Q} (\uu )$
by querying $\BBO_{\calT }$ at most once.
\end{lemm}

\begin{proof}
We need the following two propositions. Proposition~\ref{prop:allthesame}
  can be proved by mathematical induction on $d$.

\begin{prop}[\mbox{\rm Vectors not in $S$ and $Q$}]\label{prop:member}
Let $V_1=\{2,4,...,4n+4\}^{d-1}\times \{1,3,...,4n+3\}$
  and $V_k=\{2,4,...,4n+4\}^{k-2}\times\{1,3,...,4n
  +3\}\times\{2,4,...,4n+4\}^{d-k+1}$ for $2\le k\le d$.
If  $\uu\notin \cup_{k} V_k$ then $\uu $ neither appears in $S$ nor in $Q$.
\end{prop}

\begin{prop}[\mbox{\rm All the same}]\label{prop:allthesame}
Let $\calT$ and $\calT'$ two valid $(n,d)$-ToCs.
If $\uu\in \cup_{k} V_k$  and $u_i=1$ or $u_{i} = 2$
  for some $1\le i\le d$,
  then $\BBO_{S[\calT ]}(\uu)=\BBO_{S[\calT']}(\uu)$  and
  $\BBO_{Q[\calT ]}(\uu)=\BBO_{Q[\calT']}(\uu)$.
\end{prop}

We first consider two simple cases for which we  don't even
   need to query $\BBO_{\calT }$.
\begin{enumerate}
\item When $\uu\notin \cup_{k} V_k$, by Proposition \ref{prop:member},
   $\mathbb{B}_{S[\calT]}(\uu)=\mathbb{B}_{Q[\calT]}(\uu)
   =(\text{``no'',``no''})$;
\item When $\uu\in \cup_{k} V_k$  and $u_i=1$ or $u_{i} = 2$
  for some $1\le i\le d$, by Proposition \ref{prop:allthesame},
  we can compute $\BBO_{S} (\uu )$ and $\BBO_{Q} (\uu )$ from
  the valid $(n,d)$-tree in which every connector is
  generated by the identity permutation from $\{1,\dotsc ,n \}$ to $\{1,\dotsc ,n \}$.
\end{enumerate}
Now we can assume $\uu\in \cup_{k} U_k$ where
  $U_1=\{4,6,...,4n+4\}^{d-1}\times\{3,5,...,4n+3\}$
  and $U_k=\{4,6...,4n+4\}^{k-2}\times\{3,5,...,4n +3\}\times\{4,6,...,4n+4\}^{d-k+1}$
  for $2\le k\le d$.
First note that there is exactly one odd entry in $\uu $.
If $\uu \in U_{k}$ then let $\uu'$ be the string obtained from $\uu $
  by $k-1$ left-rotations.
Not the last entry of $\uu '$ is odd.
Let $\qq$ be the vector in $(\mbbJ_{n})^{d}$ where $q_{i} = u'_{i}/2$
  for $1\leq i\leq d-1$ and $q_{d} = (u'_{d}+1)/2$.
We now prove a stronger statement which implies that
  $\BBO_{S} (\uu )$ and $\BBO_{Q} (\uu )$ can be computed
  from $\BBO_{\calT } (\qq)$.

\begin{quotation}
\noindent {\em
For all $\qq \in (\mbbJ_{n})^{d}$,
  we can determine $\BBO_{S}(\uu_i)$ and $\BBO_{Q}(\uu_i)$
  from $\BBO_{\calT}(\qq)$, where
  $\uu_{k}$ is the vector in $U_{k}$ obtained from $\calF(\qq)$
   by $k-1$ right rotations, for $1\leq k\leq d$.}
\end{quotation}

If $\mathbb{B}_\calT(\qq)=\calT$, the statement is clearly true.
Otherwise, assuming $\mathbb{B}_\calT(\qq)\not=\calT$,
  we prove the statement by induction on $d$.
The base case when $d=1$ is trivial.
For $d\ge 2$,  let $\qq'=(q_2,q_3,...,q_d)$ and
  let $\vv_k$ be the vector generated from $\calF(\qq')$
  by $k-1$ right rotations, for $1\leq k\leq d-1$.
Let $\calT'$ be the subtree of $\calT$ rooted
  the $q_1$-successor of the root of $\calT $.
As $\mathbb{B}_{\calT'}(\qq')$ is contained in
  $\mathbb{B}_\calT(\qq)$,
   we can determine $\BBO_{S[\calT']}(\vv_i)$ and $\BBO_{Q[\calT']}(\vv_i)$ for
   using our inductive hypothesis,
  from which, we will show below,  we can  determine
  $\BBO_{S}(\uu_i)$ and $\BBO_{Q}(\uu_i)$.\vspace{0.025cm}

We will only prove the case for $\BBO_{S}(\uu_i)$ when $q_1$ is
  even.
All other cases are similar.\vspace{0.025cm}

Note that the first entry of $\vv_{i}$ is not 2, so for all $1\le i\le d-1$,
  $\BBO_{S[\calT']}(\vv_i)\not=(\text{``no''},a)$.
Also, for $i>1$  the last entry of $\vv_{i}$ is even,
so, $\mathbb{B}_{S[\calT']}(\vv_i)$ $\not=(a,\text{``no''})$.
Therefore,  for all $i\in[2:d-1]$, $\BBO_{S}(\uu_i)=\BBO_{S[\calT']}(\vv_i)$.
For $i=1$ or $d$, if
  $\BBO_{S[\calT']}(\vv_1)=(\text{``no''},\text{``no''})$, then
  $\BBO_{S}(\uu_i)=(\text{``no''},\text{``no''})$;
if $\BBO_{S[\calT']}(\vv_1)=(a,b)$, then $\BBO_{S}(\uu_1)=(a,2q_1)$,
  $\BBO_{S}(\uu_d)=(2q_1,b)$;
if $\BBO_{S[\calT']}(\vv_1)=(a,\text{``no''})$,
  letting the second component of $\BBO_\calT(\qq)$ be $r$,
  then $\BBO_{S}(\uu_1)=(a,2r)$ and $\BBO_{S}(\uu_d)=(\text{``no''},\text{``no''})$.
\end{proof}

\subsection{Knowledge Representation in Algorithms for \NTd and a Key Lemma}\label{}

An algorithm for \NTd  tries to learn about the connectors
  in $\calT^{*} $ by repeatedly querying its leaves.
To capture its intermediate knowledge about this $\calT^{*}$,
  we introduce a notion of partial connectors.\vspace{0.01cm}

Let $\ssigma = [\sigma (1),...,\sigma (k)]$ be an array
  of distinct elements from $\{0,1,...n \}$.
Then, $\ssigma$ defines a string
  $S_{\sigma(1)}\circ...\circ S_{\sigma (k)}$,
   referred to as a {\em
   connecting segment}.
Recall $S_0 = 2,S_1=3\circ 4,...S_n=(2n+1)\circ (2n+2)$.
A {\em partial connector} over $\mathbb{J}_{n}$ is
  then a set $\calC $ of connecting segments such that
  each $j \in \mathbb{J}_{n} $ is contained in exactly one
  segment in $\calC $ and 2 is the first element of the segment containing
  it.
If $\calC$ has $n+1$ segments,
  that is, $\calC=\{2,3\circ 4, ..., (2n+1)\circ (2n+2)\}$, then $\calC$
  is called an {\em empty connector}.
We say a  connector $C$ is {\em consistent} with a
  partial connector $\calC$ if every segment in $\calC$ is a substring of $C$.\vspace{0.02cm}

Let $r[\calC]$ be the last symbol of the segment in $\calC$
  that starts with 2.
Let $L[\calC]$ and $R[\calC]$, respectively,
  be the set of first and the last symbols of other segments in $\calC $.
So, $r[\calC] \in \mbbF_{n}\cup \{2 \}$, $L[\calC]\subset
  \mbbO_n$, and $R[\calC]\subset \mbbF_n$.
Also, $|\hspace{0.04cm}L[\calC]\hspace{0.04cm}|=|\hspace{0.04cm}R[\calC]\hspace{0.04cm}|$.
If $2\neq r[\calC]$, we use $\phi_{\calC } (2)$ to denote
  its right neighbor.
Note that each $s \in \mbbJ_{n} - L[\calC ]\cup R[\calC ] \cup
  \{r[\calC ],2 \}$ has two neighbors in $\calC $.
If $s$ is even, we will use $\phi_{\calC } (s)$ to denote its right
   neighbor and if $s$ is odd, we use $\phi_{\calC } (s)$ to denote
   its left  neighbor.\vspace{-0.1cm}
\begin{quote}
Initially, the knowledge of an algorithm for \NTd
  can be viewed as a tree $\calT $ of empty con\-nectors.
At each round, the algorithm chooses a query point $\qq$ and
  asks for $\BBO_{\calT^{*}} (\qq)$, which may connect
  some segments in the partial connectors.
So $\calT $ is updated.
The algorithm succeeds when every partial connector becomes
  a connector and $\calT $ grows into  $\calT^{*}$.
\end{quote}

So, at intermediate steps, the knowledge of the algorithm
   can be expressed by a tree $\calT $ of partial connectors.

\begin{defi}[\mbox{\rm Valid Tree of Partial
Connectors}]\label{defi:ToPCSplit}
An $(n,d)$-{\rm ToPC} $\calT $ is a tree $T_{n,d}$ in which each internal node
  $v$ is associated with a partial connector $\calC _{v}$ over $\mbbJ_{n}$.

$\calT $ is a valid $(n,d)$-{\rm ToPC} if 
  for each internal node $v\in T_{n,d}$ whose children are not leaves,
  its partial connector $\calC_{v}$ at $v$ satisfies the
  following condition:
For each pair $s,t\in \mbbJ_{n}$ with
      $\phi_{\calC_{v}} (s) =t$,
      the tree $\calT_{s}$ rooted at the $s^{th}$-successor $v_{s}$ and
      the tree $\calT_{t}$ rooted at the $t^{th}$-successor $v_{t}$ of $v$ are both
      valid ToCs, and $\name{\tail{v_{s}}}$ in $\calT_{s}$ and
      $\name{\tail{v_{t}}}$ in $\calT_{t}$
      are the same.\vspace{0.015cm}
\end{defi}

A valid $(n,d)$-ToC $\calT^{*}$ is {\em consistent}
   with a valid $(n,d)$-ToPC $\calT$,
   denoted by $\calT \models \calT^{*}$,
   if for every internal node, its connector in $\calT^{*}$ is
   consistent with its (partial) connector in $\calT$.

A partial connector
 $\calC$ is a {\em  $\beta$-partial connector} for $0<\beta<1$
  if the number of segments in $\calC$~is at least $(1-\beta) n+1$.
To simplify our proof, we will relax
   our oracle $\BBO_{\calT^* }$ to sometime provide
   more information to the algorithm than being asked
   so that the $\calT $ it maintains always satisfies the
  conditions of the following definition:

\begin{defi}[\mbox{\rm Valid $(n,d,\beta )$-ToPC}]\label{defi:ToPC}
A valid $(n,d)$-ToPC $\calT $ is a {\rm valid $(n,d,\beta )$-ToPC} 
  if its root has a $\beta  $-partial connector,
Moreover, for each internal node $v\in T_{n,d}$ whose children are not leaves,
  if the partial connector $\calC_{v}$ at $v$
  is a $\beta $-partial connector, then it satisfies the
   following condition:
The $s^{th}$-successor of $v$,
   for each $s \in L[\calC ]\cup R[\calC ] \cup \{r[\calC ]\}$,
   has a $\beta$-partial connector. \vspace{0.015cm}
\end{defi}

Key to our analysis is Lemma \ref{lem:key} below, stating
    that every valid $(n,d,\beta)$-ToPC has
    a large number of consistent valid $(n,d)$-ToCs, and moreover,
    the names of the tails of these ToCs are nearly-uniformly distributed.
Let $\calF [\calT ] = \{\hspace{0.04cm}\name{\tail{\calT^{*}}}
   \big| \hspace{0.1cm} \calT \models
  \calT^{*}\hspace{0.04cm}  \}.$ Also, for each $\pp\in (\mbbF_{n})^{d}$,
let $N[\calT,\pp] =  |\hspace{0.05cm}\{\hspace{0.04cm}\calT^{*}
  \hspace{0.06cm} \big| \hspace{0.1cm}
  \calT \models \calT^{*}  \ \text{and}\ \name{\tail{\calT^{*}}} = \pp\hspace{0.04cm}  \}
  \hspace{0.05cm} |$.

\begin{lemm}[\mbox{\rm Key Lemma}]\label{lem:key}
For $d\geq 1$ and $\beta\in [0,24^{-d}]$,
  $|\hspace{0.04cm} \calF[\calT]\hspace{0.01cm} |\ge
(\hspace{0.02cm}(1-\beta)n\hspace{0.02cm} )^d$
for each valid $(n,d,\beta)$-ToPC $\calT$.
Let $\alpha_1(\beta)=1$.
Then, for all $\pp_1,\pp_2\in \calF[\calT]$,
\begin{equation}\label{eqn:key}
\frac{1}{\alpha_d(\beta)}\hspace{0.05cm}\le\hspace{0.05cm} \frac{N[\calT,\pp_1]}{N[\calT,\pp_2]}\hspace{0.05cm}\le
\hspace{0.05cm}\alpha_d(\beta),
\quad \mbox{ where } \quad
\alpha_d(\beta)=\frac{\left(\alpha_{d-1}\left(\beta\right)\right)^7}{ (2\left(1-\beta\right)^{d-1}-1 )^3},
\mbox{ for $d \geq 2$}.\vspace{0.1cm}
\end{equation}
\end{lemm}

\begin{proof}
When $d=1$, let $\calC $ be the only partial connector in $\calT $.
Clearly, $\calF[\calT]=R[\calC]$.
Thus, in this case the lemma is true.
We will also use this case as the base of the induction below.
When $d\ge 2$,  let $\calC$ be the partial connector of
  the root.
For each $k\in \mbbJ_n$, let $\calT_k$ be the subtree
  of  the $k^{{th}}$-successor of the root.
Below, we will prove by induction on $d$ that (\ref{eqn:key})
  and (**)
$\calF[\calT]=\cup_{k\in R[\calC]}
  (k \circ \calF[\calT_{k}]\hspace{0.06cm})$
  are true for all $d$.
Note that (**) and
  the first condition of Definition \ref{defi:ToPC} imply that
  $|\hspace{0.04cm} \calF[\calT]\hspace{0.01cm} |\ge
(\hspace{0.02cm}(1-\beta)n\hspace{0.02cm} )^d.$\vspace{0.02cm}

Let $\calC=\{\hspace{0.04cm}Y_0,Y_1,Y_2,...,Y_m
  \hspace{0.04cm}\}$
  be the $\beta$-partial connector at
  the root of $\calT$; assume $Y_0$ is the
  segment starting with $2$.
We use $r_i$ and $t_{i}$, respectively,  to denote the ending and
  starting symbols of $Y_i$.
For each $k\in \{\hspace{0.04cm}r_0, ...,r_m,t_1,...,t_m
\hspace{0.04cm}\}$,
   let $\calT_k$ denote the $(n,d-1,\beta)$-ToPC at the $k^{{th}}$-successor of the root.
For each pair $(i,j)\in [0:m]\times [1:m]$ with $i\not= j$, we
define
\begin{eqnarray*}
&N_{i,j}=\sum_{\pp\in \calF[\calT_{r_i}]\cap \calF[\calT_{t_j}]} N[\calT_{r_i},\pp]\cdot N[\calT_{t_j},\pp].
\end{eqnarray*}
Inductively, (\ref{eqn:key}) and (**) hold for all $d'<d$.
As a result, we have $|\hspace{0.04cm}\calF[\calT_k]\hspace{0.04cm}| \geq ((1-\beta)n)^{d-1}$ for
  every $k\in \{\hspace{0.03cm}r_0, ...r_m,t_1,...t_m
  \hspace{0.03cm}\}$. Thus,\vspace{0.08cm}
$$|\hspace{0.04cm}\calF[\calT_{r_i}]\cap
\calF[\calT_{t_j}]\hspace{0.04cm}| =
|\hspace{0.04cm}\calF[\calT_{r_i}]\hspace{0.04cm}| +
|\hspace{0.04cm}\calF[\calT_{t_j}]\hspace{0.04cm}|-
|\hspace{0.04cm}\calF[\calT_{r_i}]\cup \calF[\calT_{t_j}]\hspace{0.04cm}|
\ge
(2(1-\beta)^{d-1}-1)\hspace{0.04cm}n^{d-1}>0,\vspace{0.08cm}
$$
because $\beta\leq 24^{-d}$.
By the inductive hypothesis, we have $N_{i,j}>0$, for all $(i,j)\in
  [0:m]\times [1:m]$ with $i\not= j$.\vspace{0.02cm}

To show (**), it suffices to prove that
  $N[\calT,\pp]>0$ if and only if
  $\pp \in \cup_{k\in R[\calC]} (k \circ \calF[\calT_{k}]\hspace{0.06cm}).$
\vspace{0.015cm}Clearly,  $N[\calT,\pp]= 0 $ for
   $\pp \not\in \cup_{k\in R[\calC]} (k \circ \calF[\calT_{k}]\hspace{0.06cm}).$
So, let us consider
  $\pp \in \cup_{k\in R[\calC]} (k \circ \calF[\calT_{k}]\hspace{0.06cm}).$
Since $p_1\in R[\calC]$, \vspace{0.015cm}WLOG, assume $p_1=r_m$.
We use $\calP$ to denote the set of permutations
  $s_0s_1...s_{m-1}$ over $[0:m-1]$ with $s_0=0$. Then\vspace{0.04cm}
\begin{eqnarray*}\label{eq:total}
&N[\calT,\pp]=\sum_{s_0s_1...s_{m-1}\in \calP}\left( \left( \prod_{i=0}^{m-2} N_{s_i,s_{i+1}} \right) \cdot
N_{s_{m-1},m} \cdot N\big[\calT_{r_m}, (p_2,p_3,...,p_d)\big]\hspace{0.03cm} \right).&
\end{eqnarray*}
By the inductive hypothesis, every item in
   the summation above is positive. So $N[\calT,\pp]>0$ and
   (**) holds for $d$.

Next, to prove (\ref{eqn:key}), consider $\pp_{1} \in \calF [\calT]$
  and $\pp_{2} \in \calF [\calT]$.
There are two basic cases.
When $p_{1,1}=p_{2,1}$,
  Eqn. (\ref{eqn:key}) follows directly from
  (\ref{eq:total}) and the inductive hypothesis.
When $p_{1,1}\neq p_{2,1}$, without loss of generality,
  we assume $p_{1,1}=r_{m}$ and $p_{2,1}=r_{m-1}$.\vspace{0.015cm}

Let $\calP_1$ denote the set of permutations
   over $\{0,1,...,m-2,m-1\}$ with $s_0=0$
    and $\calP_2$ denote the set of permutations
   over $\{0,1,...,m-2,m\}$ with $s_0=0$.
For  $P=s_0s_1...s_{m-1} \in \calP_{1}$ ,
  let $\Pi (P)$ be the
  permutation obtained from $P$ by replacing $m-1$ by $m$.
Clearly $\Pi$ is a bijection from $\calP_{1}$ to $\calP_2$.
We can write
   $N[\calT,\pp_1]$ and $N[\calT,\pp_2]$ as two summations:
\begin{eqnarray*}
&N[\calT,\pp_1]=\sum_{P\in \calP_1} N_{1} (P),\ \ \ \ \text{and}\ \ \ \
N[\calT,\pp_2]=\sum_{P\in \calP_1} N_{2} (\Pi(P)),&
\end{eqnarray*}
where $N_{1} (P)$ and $N_{2} (\Pi (P))$ are
   given by similar terms as in (\ref{eq:total}).\vspace{0.015cm}

We now prove for every $P\in \calP_1$,
  $(N_{1} (P)/N_{2} (\Pi (P))\le \alpha_d(\beta)$.
 Let $P=s_0s_1...s_{m-1}$ where $s_k = m-1$ for some
  $1\le k\le m-1$.
If $k<m-1$, then we expand  $N_{1} (P)$ and $N_{2}(\Pi (P))$  as:
\begin{eqnarray*}
\frac{N_{1} (P)}{N_{2} (\Pi (P))}=
  \frac{N_{s_{k-1},m-1}\cdot N_{m-1,s_{k+1}} \cdot N_{s_{m-1},m} \cdot
N[\calT_{r_m},(p_{1,2},p_{1,3},...,p_{1,d})]}{N_{s_{k-1},m}\cdot N_{m,s_{k+1}} \cdot N_{s_{m-1},m-1} \cdot
N[\calT_{r_{m-1}},(p_{2,2},p_{2,3},...,p_{2,d})]}.\vspace{0.03cm}
\end{eqnarray*}
It then follows from the application of our inductive hypothesis
  to the straightforward expansion of terms $N_{i,j}$
 that $N_{1} (P)/N_{2} (\Pi (P))\le \alpha_d(\beta)$.\vspace{0.02cm}

Similarly, we can establish the same bound for the case when $k=m-1$.
\end{proof}

\subsection{The Randomized Query Complexity of \NTd} \label{sub:proof}

By querying every leaf, one can solve any
  instance of \NTd with $n^{d}$ queries.
Below, we prove Theorem \ref{thm:NT} by showing
  $\text{\sf RQ}_{\text{\sf NT}}^d(n) =(\Omega (n))^{d}$.
We first relax $\BBO_{\calT^{*}}$ by extending it to
  $(\mbbJ_n)^m$ for $m\in[1: d]$.

\begin{defi}[\mbox{\rm Relaxation of $\BBO_{\calT^{*}}$}]
Suppose $\calT^{*}$ is a valid $(n,d)$-ToC and $\qq\in (\mbbJ_{n})^{m}$.
Let $v$ be the node with  $\name{v} = q_{1}q_2...q_m$.
Let $\qq' = \name{\tail{v}}\in (\mathbb{J}_n)^d$ \emph{(}in tree $\calT^{*} $\emph{)}.
Then, $\BBO_{\calT^{*}} (\qq) = \BBO_{\calT^{*}} (\qq')$.
\end{defi}

\begin{figure}[!h]
\rule{\textwidth}{1pt}\vspace{0.04cm}

\textbf{\ \ Query-and-Update$(\calT,\qq)$, where $\qq\in (\mbbJ_n)^d$}

\vspace{-0.18cm}\rule{\textwidth}{1pt}\vspace{0.16cm}

\begin{tabular}{@{\hspace{0.1cm}}r@{\hspace{0.2cm}}p{\textwidth}}
0\hspace{0.05cm}: & \ \ \textbf{if} $\calT$ has complete
  information of $\qq$ \textbf{then} return;\\ [0.5ex]

\ \ 1\hspace{0.05cm}: & \ \ \textbf{if} \hspace{0.04cm}$\exists\ 0\le i\le d-1:
\big|\hspace{0.02cm}R[\calC_i]\hspace{0.04cm}
\big|= (1-\beta_d)n$ \textbf{then} \\[0.5ex]

2\hspace{0.05cm}: & \ \ \ \ \ \ set $m$ be the smallest of such $i$ ($m\in [0:d-1]$) \\[0.5ex]

3\hspace{0.05cm}: & \ \ \textbf{else} set $m=d$\\[0.5ex]

4\hspace{0.05cm}: & \ \ \textbf{if} $m=0$ \textbf{then} set $\calT=\calT^*$\
\hspace{0.08cm}\textbf{\{}\hspace{0.05cm}and $I=1$\hspace{0.05cm}\textbf{\}}\\[0.5ex]

5\hspace{0.05cm}: & \ \ \textbf{else} \textbf{Update}$\hspace{0.06cm}(\calT,(q_1,q_2,...,q_m),m)$\\[0.65ex]
\end{tabular}
\rule{\textwidth}{1pt}\vspace{0.035cm}

\textbf{\ \ Update$(\calT ,\qq,m)$,
   where $\qq\in (\mbbJ_n)^m$ and $1\le m\le d$}

\vspace{-0.18cm}\rule{\textwidth}{1pt}\vspace{0.16cm}

\begin{tabular}{@{\hspace{0.1cm}}r@{\hspace{0.2cm}}p{\textwidth}}
\ \ 6\hspace{0.05cm}: & \ \ fetch $\mathbb{B}_{\calT^*}(\qq)$\ \hspace{0.08cm}\textbf{\{}\hspace{0.05cm}set
$A_m=A_m+1$, $\calB_m[A_m]=0$ and $\calB_{m,k}[A_m]=0$\hspace{0.05cm}\textbf{\}}\\[0.5ex]

7\hspace{0.05cm}: & \ \ \textbf{if}
$\mathbb{B}_{\calT^*}(\qq)=\calT^*$ \textbf{then}
 set $\calT=\calT^*$ \hspace{0.04cm}\textbf{\{}\hspace{0.07cm}set $\calB_m[
A_m]=1$\hspace{0.05cm}\textbf{\}}\\[0.5ex]

8\hspace{0.05cm}: & \ \
\textbf{else} [\hspace{0.08cm}let $d-m \le h\le d-1$ and $r\in \mbbJ_n$
  be the first and second
components of $\mathbb{B}_{\calT^*}(\qq)$\hspace{0.06cm}]\\[0.5ex]

9\hspace{0.05cm}: & \ \ \ \ \ \ set $m' = d-h-1$ \\ [0.5ex]

10\hspace{0.05cm}: & \ \ \ \ \ \ $\exists\ Y_1,Y_2\in \calC_{m'}$: $\{\hspace{0.05cm}\text{the ending symbol of $Y_1$,
the starting symbol of $Y_2$}\hspace{0.05cm}\}=\{\hspace{0.04cm}q_{m'+1},r\hspace{0.04cm}\}$\\ [0.5ex]

11\hspace{0.05cm}: & \ \ \ \ \ \ replace $Y_1$ and $Y_2$ in $\calC_{m'}$ by the concatenation of $Y_1$ and $Y_2$
\hspace{0.04cm}\textbf{\{}\hspace{0.05cm}set $\calB_{m,m'}[A_m]=1$\hspace{0.05cm}\textbf{\}}\\[0.5ex]

12\hspace{0.05cm}: & \ \ \ \ \ \ let $\calT'$ and $\calT''$ be the
third and fourth components of $\mathbb{B}_{\calT^*}(\qq)$ \\[0.5ex]

13\hspace{0.05cm}: & \ \ \ \ \ \ replace the subtree of $\calT$ rooted at $u_{m'+1}$ with $\calT'$;\\[0.5ex]

14\hspace{0.05cm}: & \ \ \ \ \ \  replace the subtree of $\calT$ rooted at the $r$-successor of $u_{m'}$ with
$\calT''$\\[0.65ex]
\end{tabular}
\vspace{-0.2cm}\rule{\textwidth}{1pt}\vspace{0.2cm}
\caption{
}\label{fig:query}
\end{figure}

\begin{proof}[{\bf Proof} \mbox{\rm (Theorem \ref{thm:NT})}]
To apply Yao's Minimax Principle \cite{Yao77}, we
  consider the distribution $\calD$ in which each
   valid $(n,d)$-ToC $\calT^{*}$ is chosen with the same probability.
We will prove that the expected query complexity of any deterministic
  algorithm $\calA $ for \NTd over $\calD$ is $(\Omega (n))^{d}$.
Let $\beta_d=24^{-d}$.\vspace{0.02cm}

Suppose, at a particular step, the current knowledge of $\calA $ can be
  expressed by a valid $(n,d,\beta_{d})$-ToPC $\calT $, which is
  clearly true
  initially,  and $\calA $ wants
  to query $\qq \in (\mbbJ_{n})^{d}$.
Let $u_0$ be the root of $\calT$ and $u_i$ be the node with
  $\name{u_{i}} = q_1...q_i$.
Let $\calC_i$ be the partial connector at $u_i$ in $\calT$
  and $\calT_i$ be the subtree of $\calT$ of $u_i$.
There are two cases
(1) $\forall i \in [0,d-1]$, $\calC_i$ is a partial connector and
  $q_{i+1}\in L[\calC_i]\cup R[\calC_i]\cup \{r[\calC_i]\}$.
(2) otherwise.
From the definition of $\BBO_{\calT^{*}}$,
  we can show that in case (2),   $\BBO_{\calT^{*}} (\qq)$ can be
  answered based on $\calT $ only.
So, WLOG, we assume $\calA $ is smart and never asks unnecessary queries.\vspace{0.015cm}

In case (1), because $\calT $ is a $(n,d,\beta_{d})$-ToPC,
   $\calC_i$ is a $\beta_{d}$-partial  connector for all $i \in
  [0,d-1]$.
Let $h=\height{\head{\qq}}$.
If $h=d$, then $\calA $ gets $\calT^{*}$.
Otherwise, the knowledge gained by querying $\BBO_{\calT^{*}} (\qq)$
   connects two segments in $\calC_{d-h-1}$
   and  replaces the two involved subtrees by the corresponding
   ones in $\BBO_{\calT^{*}} (\qq)$.
The resulting tree $\calT $, however, may no longer be a
  $(n,d,\beta_{d})$-ToPC, if
  $|R [\calC_{d-h-1}]| = (1-\beta_{d})n$ before the query.
We will relax $\BBO_{\calT^{*}}$ to provide $\calA$ more information
  to ensure that the resulting $\calT $ remains a valid $(n,d,\beta_{d})$-ToPC.
To this end, we  consider two subcases:
Case (1.a): if $\forall\ i\in[0:d-1]$,  $|R [\calC_{i}]| > (1-\beta_{d})n$,
  then $\calA$ receives $\BBO_{\calT^{*}} (\qq)$ as it requested.
Case (1.b): if $\exists\ i\in[0:d-1]$ such that
  $|R [\calC_{i}]| = (1-\beta_{d})n$, then
   let $m = \min\hspace{0.04cm} \{\hspace{0.04cm}i\hspace{-0.03cm}:\hspace{-0.03cm}  |R [\calC_{i}]| =
   (1-\beta_{d})n\hspace{0.04cm}\}$.
Let $\qq' = (q_{1},...,q_{m})$. Instead of getting $\BBO_{\calT^{*}} (\qq)$,
   $\calA$ gets $\BBO_{\calT^{*}} (\qq')$.
In this way, the resulting $\calT $ remains a valid $(n,d,\beta_{d})$-ToPC.
Details of the query-and-update procedure can be found in
  Figure~\ref{fig:query}.\vspace{0.015cm}

We introduce some ``analysis variables'' to aid our analysis.
These variables include:\hspace{-0.05cm}
(1) $I\in \{ 0,1 \}$:
Initially, $I=0$.
If $m = 0$ in case (1.b), then we set $I = 1$.
(2) For each $m\in[1:d]$,  $A_m\in \mathbb{Z}$,
    and~a set of binary sequences
    $\calB_{m}[...]$ and $\calB_{m,k}[...]$, $\forall k\in[0:m-1]$.
Initially, $A_m=0$, and $\calB_{m}$, $\calB_{m,k}$ are empty.
Each time in case (1.b) when $m>0$, we increase  $A_{m}$ by $1$;
in case (1.a), we increase $A_{d}$~by~$1$.
To unify the discussion below, if we have case (1.a),
  let $m=d$ and $\qq' =\qq$.
If $\BBO_{\calT^{*}} (\qq') = \calT^{*}$, we set
  $\calB_{m}[A_{m}]=1$ and $\calB_{m,k}[A_{m}]=0$,  $\forall k\in [0: m-1]$.
Otherwise, if the first component of $\BBO_{\calT^*}(\qq')$ is
   $d-l$, for $l\in [1:m]$, then set $\calB_{m,l-1}[A_{m}]=1$,
    $\calB_{m}[A_{m}]=0$ and $\calB_{m,k}[A_{m}]=0$ for all $0\le k\not=l-1\le m-1$.\vspace{0.025cm}

Let  $M_d=(\beta_d n/2)^d$.
Given a random valid $(n,d)$-ToC $\calT^{*} $,
  if $\calA $ stops before making $M_{d}$ queries,
  let $\{ I,A_m,\calB_m,\calB_{m,k} \}$
  be the set of analysis variables  assigned when $\calA $ stops;
otherwise, $\{ I,A_m,\calB_m,\calB_{m,k} \}$
  is assigned after  $\calA $ makes exactly $M_{d}$
  queries.
Let $M_i=(\beta_d n/2)^i$.
We define a set of binary strings
  $\{\hspace{0.06cm}\overline{\calB}_m[1...M_m],\overline{\calB}_{m,k}[1...M_m],
 \hspace{0.05cm} 1\le m\le d,\hspace{0.05cm} 0\le k\le m-1\hspace{0.06cm}\}$ from $\calB_m$ and $\calB_{m,k}$:
For every $1\le i\le M_m$,
(I) $\overline{\calB}_m[i]=\calB_m[i]$ for $i\le \min (A_m,M_{m})$
  and $\overline{\calB}_m[i]=0$ for $A_{m} < i \leq M_{m}$;
(II) $\overline{\calB}_{m,k}[i]=\calB_{m,k}[i]$ for $i\le \min
      (A_m,M_{m})$ and  $\overline{\calB}_{m,k}[i]=0$ $A_{m} < i \leq M_{m}$.\vspace{0.025cm}

Let $\EventTrue{A}$ denote that an event $A$ is true.
Let $\notyetfound{\calT^{*} }$ be the event that
  $\calA$ hasn't found the tail of $\calT^{*} $
  after making $M_{d}$ queries.
Let  ${B}_m$, ${B}_{m,k}$,  $\overline{B}_m$ and $\overline{B}_{m,k}$
  denote the number of $1$'s in
${\calB}_{m}$,  ${\calB}_{m,k}$,
$\overline{\calB}_{m}$ and
  $\overline{\calB}_{m,k}$, respectively.
Then,
 $\EventTrue{\notyetfound{\calT^{*} }}$ if and only if
 $[I=0  \mbox{ and}$ $B_m=0, \forall  m\in [1:d]]$.
The theorem directly follows from Lemmas \ref{lem:logic} below.
\end{proof}

\begin{lemm}\label{lem:logic}
Let $A$ denote the following event,
\begin{eqnarray*}
 A = \left(\hspace{0.05cm}\overline{B}_m=0 \mbox{ and }
  \overline{B}_{m,k}\le \frac{16\cdot M_m}{n^{m-k-1}} \mbox{ and }
  \overline{B}_{m,m-1}\le M_m, \forall\ m\in [1:d], k\in[0,m-2]\hspace{0.05cm}\right).
\end{eqnarray*}
then \mbox{\rm \textbf{(E.1)}}
  $\EventTrue{A}$ implies $\EventTrue{ \notyetfound{\calT
  } }$ and \mbox{\rm \textbf{(E.2)}}
 $\mbox{\sf Pr}_{\calD }\left[A \right] \geq 1/2$.
\end{lemm}

\begin{proof}[{\bf Proof} \mbox{\rm (of Lemma
\ref{lem:logic})}:]
To prove \textbf{(E.1)}, we use the following inequalities
  that follow from the definition of our analysis variables.
\begin{equation}\label{eqn:fact}
\text{1)}\ \ A_m\le \frac{1}{\beta_d n}\sum_{i=m+1}^d B_{i,m},\ \text{for all $1\le m\le d-1$};\ \ \text{and\ \ 2)}\
\ I=1\ \Longrightarrow \ \sum_{i=1}^d B_{i,0}\ge \beta_d n.
\end{equation}


Recall that $[\hspace{0.05cm}\notyetfound{\calT }\hspace{0.03cm}] =
[\hspace{0.05cm}I = 0 \mbox{
and } B_{m} = 0, \ \forall m \in [1:d]\hspace{0.05cm}].
$\vspace{0.015cm}

To prove \textbf{(E.1)}, it suffices to show that $\EventTrue{A} \Rightarrow
 \EventTrue{\hspace{0.04cm}I = 0\hspace{0.04cm}}$ and
 $\EventTrue{A} \Rightarrow \EventTrue{\hspace{0.04cm}B_{m} = 0,
 \ \forall\ m \in [1:d]\hspace{0.04cm}}$.
We use $\EventTrue{A} \Rightarrow \EventTrue{B}$ to denote
  if event $A$ is true then event $B$ is true.
It follows immediately from the definitions of $B_{m}$ and $\overline{B}_m$,
  that if   $A_{m} \leq M_{m}$, then $B_{m} = \overline{B}_m$.
So, we first inductively prove that $\EventTrue{A} \Rightarrow
  \EventTrue{\hspace{0.04cm}A_{d-m} \leq M_{d-m},
  \forall\ m\in [0:d-1]\hspace{0.04cm}}$.
The base case when $m=0$ is trivial, since $A_d$ is at most $M_d$,
  the total number of queries.\vspace{0.01cm}

We now consider $m\ge 1$, and assume inductively,
   that $A_i\le M_i$ for all $i\in[d-m+1: d]$.
Consequently, for all $i\in[d-m+1:d]$ and $j\in [0,i-1]$, $\overline{B}_i=B_i$ and
  $\overline{B}_{i,j}=B_{i,j}$.
By Eqn. (\ref{eqn:fact}), we have
\begin{eqnarray*}
A_{d-m} \hspace{-0.5cm}&&\le
  \sum_{i=d-m+1}^{d} \frac{B_{i,d-m}}{\beta_d n} = \sum_{i=d-m+1}^{d}
   \frac{\overline{B}_{i,d-m}}{\beta_d n}
  \le \frac{1}{\beta_d n} \left(M_{d-m+1}+\sum_{i=d-m+2}^d
   \left( \frac{16\cdot M_{i}}{n^{i-d+m-1}} \right)\right)\\[1.3ex]
   &&\le M_{d-m}\left(\frac{1}{2}+8\sum_{i=d-m+2}^d
   \left(\frac{\beta_d}{2}\right)^{i-d+m-1}\right)
\le M_{d-m}\left(\frac{1}{2}+8\cdot \frac{\beta_d}{2}\cdot 2\right)< M_{d-m}.
\end{eqnarray*}
Thus, $\EventTrue{A} \Rightarrow \EventTrue{\hspace{0.04cm}B_{m}
  = 0, \ \forall \ m\in[1:d]\hspace{0.04cm}}$.
Now we prove $\EventTrue{A}$ implies $\EventTrue{\hspace{0.04cm}I=0\hspace{0.04cm}}$.\vspace{0.025cm}

Consider the partial connector $\calC $ at the root.
We have,
\begin{eqnarray*}
&\Big[\hspace{0.08cm}B_{m} = 0, \ \forall \ m\in[1:d] \mbox{ and }
\sum_{m=1}^{d}B_{m,0}
< \beta_d n\hspace{0.08cm}\Big] \Longrightarrow \big[\hspace{0.05cm}
|\hspace{0.04cm}R [\calC ]\hspace{0.04cm} | > (1-\beta_d )n
\hspace{0.05cm}\big]
\Longrightarrow    \EventTrue{\hspace{0.04cm}I=0\hspace{0.04cm}}.&
\end{eqnarray*}
So it suffices to show $\EventTrue{A}$ implies
   $\big[\sum_{m=1}^{d}B_{m,0}< \beta_d n\big]$.
Assume $\EventTrue{A}$, then
\begin{equation*}
\sum_{i=m}^d B_{m,0}= \overline{B}_{1,0}+\sum_{m=2}^d
\overline{B}_{m,0}
  \le M_1+\sum_{m=2}^d \frac{16\cdot M_m}{n^{m-1}}=\beta_d n
  \left(\hspace{0.06cm} \frac{1}{2} +8\sum_{m=2}^d \left( \frac{\beta_d}{2}
\right)^{m-1}\right)<\beta_d n.
\end{equation*}

The first equation follows from $\EventTrue{A} \Rightarrow
\EventTrue{A_{d-m} \leq M_{d-m}, \forall m\in [0:d-1]}$ and
  the first inequality uses $\overline{B}_{m,m-1}\le M_m$ for all
  $m\in [1:d]$.
Finally, to prove \textbf{(E.2)},
\begin{eqnarray*}\label{eq:equiv1}
\mbox{\sf Pr}_{\calD }\left[A\right] & = &
  \mbox{\sf Pr}_{\calD }\left[\hspace{0.09cm}\overline{B}_m=0,
  \overline{B}_{m,k}\le \frac{16\cdot M_m}{n^{m-k-1}},
  \overline{B}_{m,m-1}\le M_m,\forall\ m\in [1:d], k\in [0:m-2]\hspace{0.09cm}\right]\\
   & \ge & 1 - \left(\hspace{0.06cm}\sum_{m=1}^{d}
   \mbox{\sf Pr}_{\calD }\left[ \overline{B}_m > 0\right] +
  \sum_{m=1}^{d}\sum_{k=1}^{m-2}  \mbox{\sf Pr}_{\calD }\left[\hspace{0.04cm}
  \overline{B}_{m,k} >\frac{16\cdot M_{m}}{n^{m-k-1}}
  \hspace{0.04cm}\right]
\hspace{0.06cm}\right) \ge \frac{1}{2}. \end{eqnarray*}
The last inequality follows from Lemma \ref{lem:prob}.
\end{proof}

As $\calT $ is chosen randomly from valid $(n,d)$-ToCs,
  $\overline{\calB}_m$ and $\overline{\calB}_{m,k}$ are random
  binary strings from a distribution
  defined by the deterministic algorithm $\calA $.
To assist the analysis of these random binary strings, we
  introduce the following definition.\vspace{0.03cm}

\begin{defi}[\mbox{\rm $c$-Biased Distributions}]
Suppose we have a probabilistic distribution over
$\{\hspace{0.03cm}0,1\hspace{0.03cm}\}^m$.
For every binary string $S$ of length at most $m$,
we define $$
U_S=\Big\{\hspace{0.05cm}\text{$S' \in
\{\hspace{0.03cm}0,1\hspace{0.03cm}\}^m \hspace{0.03cm}
  \big| \hspace{0.05cm}
   S$ is a prefix of $S'$}\hspace{0.05cm}\Big\}.
$$
For $0\le c\le 1$, the distribution is said to be {\em  $c$-biased} if we
  have
  $\mbox{\sf Pr}[U_{1}]\le c$ and
   $\mbox{\sf Pr}[U_{S\circ 1}]\le c\cdot \mbox{\sf Pr}[U_S]$ for every
   binary string $S$ with $1\le |S|\le m-1$.\vspace{0.05cm}
\end{defi}

As an important step in our analysis, we prove the following lemma.\vspace{0.03cm}

\begin{lemm}[\mbox{\rm Always Biased}]\label{lem:biased}
For all $1\le m\le d$,
  the distribution over
  $\overline{\calB}_m$ is $2/n^m$-biased.
Similarly, for $2\le m\le d$ and $0\le k\le  m-2$,
  the distribution over
$\overline{\calB}_{m,k}$ is $2/n^{m-k-1}$-biased.
\end{lemm}
\begin{proof}
The lemma follows from Corollary  \ref{coro:imp} below of our
  Key Lemma (\ref{lem:key}).
\end{proof}

\begin{coro}\label{coro:imp}
For $d\geq 1$ and $\beta\in [0,24^{-d}]$, let $\calT $ be a valid
  $(n,d,\beta)$-ToPC and let integer
  $N = \sum_{\pp\in \calF[\calT]}N[\calT,\pp]$ be the number of
\vspace{-0.06cm} consistent ToCs.
For $\qq\in (\mbbJ_n)^m$ where $m \in [1:d]$,
  if tree $\calT$ has no information on $\qq$, then
\begin{enumerate}
\item $(N^*/N)\le (2/n^{m})$ where $N^*=|\hspace{0.04cm}\{\hspace{0.04cm}
   \calT'\hspace{0.07cm}\big|\hspace{0.12cm}\mathbb{B}_{\calT'}(\qq)=\calT',
   \calT\models \calT'\hspace{0.04cm}\}\hspace{0.04cm}|$; and
\item $(N_k/N)\le (2/n^{m-k-1})$ where for $0\le k\le m-2$,
   $N_k$ denotes the number of consistent ToCs $\calT'$ such that the first
   component of $\mathbb{B}_{\calT'}(\qq)$ is $d-k-1$.
\end{enumerate}
\end{coro}

\begin{proof}
For each $k\in [0: m-1]$, let $W_k= \{\hspace{0.04cm}\pp\in
  \calF[\calT_k]\subset(\mbbF_n)^{d-k},\
   \text{where\ }p_i=q_{k+i},\forall i\in [1:m-k]\hspace{0.04cm}\}.$
Clearly, $|\hspace{0.03cm}W_k\hspace{0.03cm}|\le n^{d-m}$.
By Lemma \ref{lem:key}, for all $\pp_{1}$ and $\pp_{2} \in \calF
  [\calT ]$, $N[\calT,\pp_{1}]/ N[\calT,\pp_{2}] \leq  \alpha_d(\beta)$. Thus\vspace{-0.2cm}
\begin{equation*}
\frac{N^*}{N}  = \frac{\sum_{\pp\in W_0} N[\calT,\pp]}{\sum_{\pp\in\calF[\calT]}N[\calT,\pp]}
  \leq  \frac{\alpha_d(\beta)\cdot|W_{0}|}{|\calF[\calT ]|} \leq
 \frac{\alpha_d(\beta) n^{d-m}}{((1-\beta)n)^d} \leq \frac{2}{n^{m}}.
\end{equation*}

The third inequality uses Proposition~\ref{lem:usefulApproximation}.
To prove the second statement, for $k \in [0: m-2]$,
   we consider any connector $C^{*}$ over $\mathbb{J}_n$ that is consistent with
   $\calC_{k}$ and satisfies
   $\phi_{C^{*}}(Q_{k+1})\not= \text{``no''}$.
Assume $\phi_{C^*}(q_{k+1})=r$.
We use $\calT'$ to denote the subtree of $\calT$ rooted at the
$r^{th}$-successor
  of $u_k$.
Since $\calT$ has no information of $\qq$,
  both $\calT_{k+1}$ and $\calT'$ are
  $(n,d-k-1,\beta)$-ToPCs.
Then\vspace{0.1cm}
\begin{equation*}
\frac{\sum_{\pp\in W_{k+1}\cap \calF[\calT_{k+1}]\cap \calF[\calT']}N[\calT_{k+1},\pp]\cdot N[\calT',\pp]}{\sum_{\pp\in
\calF[\calT_{k+1}]\cap \calF[\calT']}N[\calT_{k+1},\pp]\cdot N[\calT',\pp]}\hspace{0.05cm}\le\hspace{0.05cm}
\frac{(\alpha_{d-k-1}(\beta))^2\cdot n^{d-m}}{(2(1-\beta)^{d-k-1}-1)\cdot n^{d-k-1}}\hspace{0.05cm}\le\hspace{0.05cm}
\frac{2}{n^{m-k-1}}. \vspace*{-0.150in}
\end{equation*}
\end{proof}

\begin{lemm}\label{lem:prob}
 For all $m\in [1:d]$ and  $k\in [0:m-2]$, we have
$$\mbox{\sf Pr}_{\calD} \big[\hspace{0.06cm}\overline{B}_m>0\hspace{0.06cm}\big]<\frac{1}{2d^2}
\ \ \ \mbox{and}\ \ \
\mbox{\sf Pr}_{\calD} \big[\hspace{0.06cm}\overline{B}_{m,k}>
   \frac{16\cdot M_m}{n^{m-k-1}}\hspace{0.06cm} \big]<\frac{1}{2d^2}.
$$\end{lemm}

\begin{proof}
We will use the following fact:
Let $\calD_{IND}^{m}$ be the distribution over $\{0,1 \}^{m}$
  where each bit of the string is chosen independently and is equal to
  $1$ with probability $c$.
For all $c$-biased distribution $\calD^m$ over $\{0,1 \}^{m}$,
  for any $1\leq k \leq m$,
$$    \mbox{\sf Pr}_{S\leftarrow \calD^{m}} \big[\hspace{0.06cm}\mbox{$S$ has at least $k$
1's}\hspace{0.06cm} \big] \leq
     \mbox{\sf Pr}_{S\leftarrow\calD_{IND}^{m}} \big[\hspace{0.06cm}\mbox{$S$ has at
     least $k$ 1's}\hspace{0.06cm} \big].
$$

By Lemma \ref{lem:biased}, $\mbox{\sf
Pr}_{\calD} [\hspace{0.04cm}\overline{B}_m>0\hspace{0.04cm} ]
  \leq 1 -  (1-2 n^{-m}  )^{M_{m}}
\leq  4 (\beta_{d}/2  )^{m} \leq 1/2d^{2}.$
The second inequality uses Propositions \ref{lem:analy} and
 \ref{pro:tayler}, and the last inequality uses $\beta_d=24^{-d}$
  and the fact $m\geq 1$.
We can apply the Chernoff
  bound \cite{CHERNOFF} and  Lemma \ref{lem:biased}
  to prove the second probability bound.
\end{proof}

\section{A Conjecture}

We conclude this paper with the following conjecture.

\begin{conj}[PLS to PPAD Conjecture]
If {\rm \textbf{PPAD}} is in {\rm \textbf{P}}, then {\rm \textbf{PLS}}
   is in {\rm \textbf{P}}.
\end{conj}

\section{Acknowledgments}\label{sec:}

We would like to thank Dan Spielman, Xiaoming Sun, and
  Xiaotie Deng for discussions that are invaluable to this work.
We, especially Shang-Hua, thank
  Dan  for strongly expressing his insightful frustration two years ago about
  the lack of a good measure-of-progress in equilibrium computation
  during our conversation about whether 2-NASH can be solved in smoothed
  polynomial time.
We may never come up with the conjecture that led us to our main result had we
  not run into Xiaoming a year ago outside the entrance of the
   subway station by
  the City University of Hong Kong.
Leaning against the wall between the supermarket and the subway station,
  Xiaoming introduced us to the work of
  Aldous's and Aaronson's on randomized and quantum local search.
It was this conversation that started our speculation
  that fixed-point computation in
  randomized query might be harder than local search due to the
  lack of a measure-of-progress.
We also thank Xiaoming for sharing his great insights on hiding
  random long paths inside $\Zdn$ in the
  lower bound argument for randomized and quantum local search when
  we were all at Tsinghua last summer.
We, especially Xi, thank Xiaotie for the  previous
  collaborations on fixed-point computation
  which proved to be very helpful to our technical work.
We also would like to thank Stan Sclaroff
  for patiently teach us the punctuation rules in English
  when writing down a conversation.


\appendix

\section{Inequalities}\label{sec:}

\begin{prop}\label{pro:tayler}
For all $\beta\ge 0$, $1-\beta\le e^{-\beta}$.
\end{prop}

\begin{prop}\label{lem:analy}
For all $0\le \beta\le 1/3$, $1-\beta\ge e^{-2\beta}$.
\end{prop}

\begin{lemm}\label{lem:usefulApproximation}
For all $d\ge 1$ and $\beta\in [0,24^{-d}]$, $\alpha_d(\beta)\le e^{2\cdot 24^{d-1}\beta}$.
\end{lemm}
\begin{proof}
We will use induction on $d$.
The base case  when $d=1$ is trivial.
We now consider the case when $d\ge 2$ and assuming inductively that
  the statement is true for all $d-1$.

By Proposition~\ref{lem:analy}, for any $\beta\in [0,24^{-d}]$, we have
\begin{eqnarray*}
\left(2\left(1-\beta\right)^{d-1}-1\right)^3&\ge&
  \left(2\left(e^{-2\beta}\right)^{d-1}-1\right)^3
  \ge \Big(2\big(1-2\beta(d-1)\big)-1\Big)^3\\
  &=& \Big(1-4\beta(d-1)\Big)^3\ge \left(e^{-8\beta\left(d-1\right)}\right)^3=e^{-24\beta(d-1)}
\end{eqnarray*}
By the inductive hypothesis, we have
\begin{equation*}
\alpha_d(\beta)\le (e^{2\cdot 24^{d-2}\beta})^7\cdot e^{24\beta(d-1)} \leq e^{2\cdot 24^{d-1}\beta},
\end{equation*}
where the last inequality follows from $14\cdot 24^{d-2}+24(d-1)\le
2\cdot 24^{d-1}$, for all $d\ge 2$.
\end{proof}

\end{document}